\pdfoutput=1
\documentclass[11pt,twoside,a4paper,pdftex,cmspaper,final,collab]{cms-tdr}

\begin{document}\cmsNoteHeader{CFT-09-007}
%
%
%

%
%
\hyphenation{env-iron-men-tal}
\hyphenation{had-ron-i-za-tion}
\hyphenation{cal-or-i-me-ter}
\hyphenation{de-vices}
%
%
\RCS$Revision: 1.34 $
\RCS$Date: 2009/11/21 17:38:28 $
\RCS$Name:  $
%
%
%

\newcommand {\etal}{\mbox{et al.}\xspace} 
\newcommand {\ie}{\mbox{i.e.}\xspace}     
\newcommand {\eg}{\mbox{e.g.}\xspace}     
\newcommand {\etc}{\mbox{etc.}\xspace}     
\newcommand {\vs}{\mbox{\sl vs.}\xspace}      
\newcommand {\mdash}{\ensuremath{\mathrm{-}}} 

\newcommand {\Lone}{Level-1\xspace} 
\newcommand {\Ltwo}{Level-2\xspace}
\newcommand {\Lthree}{Level-3\xspace}

\providecommand{\ACERMC} {\textsc{AcerMC}\xspace}
\providecommand{\ALPGEN} {{\textsc{alpgen}}\xspace}
\providecommand{\CHARYBDIS} {{\textsc{charybdis}}\xspace}
\providecommand{\CMKIN} {\textsc{cmkin}\xspace}
\providecommand{\CMSIM} {{\textsc{cmsim}}\xspace}
\providecommand{\CMSSW} {{\textsc{cmssw}}\xspace}
\providecommand{\COBRA} {{\textsc{cobra}}\xspace}
\providecommand{\COCOA} {{\textsc{cocoa}}\xspace}
\providecommand{\COMPHEP} {\textsc{CompHEP}\xspace}
\providecommand{\EVTGEN} {{\textsc{evtgen}}\xspace}
\providecommand{\FAMOS} {{\textsc{famos}}\xspace}
\providecommand{\GARCON} {\textsc{garcon}\xspace}
\providecommand{\GARFIELD} {{\textsc{garfield}}\xspace}
\providecommand{\GEANE} {{\textsc{geane}}\xspace}
\providecommand{\GEANTfour} {{\textsc{geant4}}\xspace}
\providecommand{\GEANTthree} {{\textsc{geant3}}\xspace}
\providecommand{\GEANT} {{\textsc{geant}}\xspace}
\providecommand{\HDECAY} {\textsc{hdecay}\xspace}
\providecommand{\HERWIG} {{\textsc{herwig}}\xspace}
\providecommand{\HIGLU} {{\textsc{higlu}}\xspace}
\providecommand{\HIJING} {{\textsc{hijing}}\xspace}
\providecommand{\IGUANA} {\textsc{iguana}\xspace}
\providecommand{\ISAJET} {{\textsc{isajet}}\xspace}
\providecommand{\ISAPYTHIA} {{\textsc{isapythia}}\xspace}
\providecommand{\ISASUGRA} {{\textsc{isasugra}}\xspace}
\providecommand{\ISASUSY} {{\textsc{isasusy}}\xspace}
\providecommand{\ISAWIG} {{\textsc{isawig}}\xspace}
\providecommand{\MADGRAPH} {\textsc{MadGraph}\xspace}
\providecommand{\MCATNLO} {\textsc{mc@nlo}\xspace}
\providecommand{\MCFM} {\textsc{mcfm}\xspace}
\providecommand{\MILLEPEDE} {{\textsc{millepede}}\xspace}
\providecommand{\ORCA} {{\textsc{orca}}\xspace}
\providecommand{\OSCAR} {{\textsc{oscar}}\xspace}
\providecommand{\PHOTOS} {\textsc{photos}\xspace}
\providecommand{\PROSPINO} {\textsc{prospino}\xspace}
\providecommand{\PYTHIA} {{\textsc{pythia}}\xspace}
\providecommand{\SHERPA} {{\textsc{sherpa}}\xspace}
\providecommand{\TAUOLA} {\textsc{tauola}\xspace}
\providecommand{\TOPREX} {\textsc{TopReX}\xspace}
\providecommand{\XDAQ} {{\textsc{xdaq}}\xspace}

\newcommand {\DZERO}{D\O\xspace}     


\newcommand{\de}{\ensuremath{^\circ}}
\newcommand{\ten}[1]{\ensuremath{\times \text{10}^\text{#1}}}
\newcommand{\unit}[1]{\ensuremath{\text{\,#1}}\xspace}
\newcommand{\mum}{\ensuremath{\,\mu\text{m}}\xspace}
\newcommand{\micron}{\ensuremath{\,\mu\text{m}}\xspace}
\newcommand{\cm}{\ensuremath{\,\text{cm}}\xspace}
\newcommand{\mm}{\ensuremath{\,\text{mm}}\xspace}
\newcommand{\mus}{\ensuremath{\,\mu\text{s}}\xspace}
\newcommand{\keV}{\ensuremath{\,\text{ke\hspace{-.08em}V}}\xspace}
\newcommand{\MeV}{\ensuremath{\,\text{Me\hspace{-.08em}V}}\xspace}
\newcommand{\GeV}{\ensuremath{\,\text{Ge\hspace{-.08em}V}}\xspace}
\newcommand{\TeV}{\ensuremath{\,\text{Te\hspace{-.08em}V}}\xspace}
\newcommand{\PeV}{\ensuremath{\,\text{Pe\hspace{-.08em}V}}\xspace}
\newcommand{\keVc}{\ensuremath{{\,\text{ke\hspace{-.08em}V\hspace{-0.16em}/\hspace{-0.08em}c}}}\xspace}
\newcommand{\MeVc}{\ensuremath{{\,\text{Me\hspace{-.08em}V\hspace{-0.16em}/\hspace{-0.08em}c}}}\xspace}
\newcommand{\GeVc}{\ensuremath{{\,\text{Ge\hspace{-.08em}V\hspace{-0.16em}/\hspace{-0.08em}c}}}\xspace}
\newcommand{\TeVc}{\ensuremath{{\,\text{Te\hspace{-.08em}V\hspace{-0.16em}/\hspace{-0.08em}c}}}\xspace}
\newcommand{\keVcc}{\ensuremath{{\,\text{ke\hspace{-.08em}V\hspace{-0.16em}/\hspace{-0.08em}c}^\text{2}}}\xspace}
\newcommand{\MeVcc}{\ensuremath{{\,\text{Me\hspace{-.08em}V\hspace{-0.16em}/\hspace{-0.08em}c}^\text{2}}}\xspace}
\newcommand{\GeVcc}{\ensuremath{{\,\text{Ge\hspace{-.08em}V\hspace{-0.16em}/\hspace{-0.08em}c}^\text{2}}}\xspace}
\newcommand{\TeVcc}{\ensuremath{{\,\text{Te\hspace{-.08em}V\hspace{-0.16em}/\hspace{-0.08em}c}^\text{2}}}\xspace}

\newcommand{\pbinv} {\mbox{\ensuremath{\,\text{pb}^\text{$-$1}}}\xspace}
\newcommand{\fbinv} {\mbox{\ensuremath{\,\text{fb}^\text{$-$1}}}\xspace}
\newcommand{\nbinv} {\mbox{\ensuremath{\,\text{nb}^\text{$-$1}}}\xspace}
\newcommand{\percms}{\ensuremath{\,\text{cm}^\text{$-$2}\,\text{s}^\text{$-$1}}\xspace}
\newcommand{\lumi}{\ensuremath{\mathcal{L}}\xspace}
\newcommand{\Lumi}{\ensuremath{\mathcal{L}}\xspace}
%
\newcommand{\LvLow}  {\ensuremath{\mathcal{L}=\text{10}^\text{32}\,\text{cm}^\text{$-$2}\,\text{s}^\text{$-$1}}\xspace}
\newcommand{\LLow}   {\ensuremath{\mathcal{L}=\text{10}^\text{33}\,\text{cm}^\text{$-$2}\,\text{s}^\text{$-$1}}\xspace}
\newcommand{\lowlumi}{\ensuremath{\mathcal{L}=\text{2}\times \text{10}^\text{33}\,\text{cm}^\text{$-$2}\,\text{s}^\text{$-$1}}\xspace}
\newcommand{\LMed}   {\ensuremath{\mathcal{L}=\text{2}\times \text{10}^\text{33}\,\text{cm}^\text{$-$2}\,\text{s}^\text{$-$1}}\xspace}
\newcommand{\LHigh}  {\ensuremath{\mathcal{L}=\text{10}^\text{34}\,\text{cm}^\text{$-$2}\,\text{s}^\text{$-$1}}\xspace}
\newcommand{\hilumi} {\ensuremath{\mathcal{L}=\text{10}^\text{34}\,\text{cm}^\text{$-$2}\,\text{s}^\text{$-$1}}\xspace}


\newcommand{\zp}{\ensuremath{\mathrm{Z}^\prime}\xspace}


\newcommand{\kt}{\ensuremath{k_{\mathrm{T}}}\xspace}
\newcommand{\BC}{\ensuremath{{B_{\mathrm{c}}}}\xspace}
\newcommand{\bbarc}{\ensuremath{{\overline{b}c}}\xspace}
\newcommand{\bbbar}{\ensuremath{{b\overline{b}}}\xspace}
\newcommand{\ccbar}{\ensuremath{{c\overline{c}}}\xspace}
\newcommand{\JPsi}{\ensuremath{{J}/\psi}\xspace}
\newcommand{\bspsiphi}{\ensuremath{B_s \to \JPsi\, \phi}\xspace}
\newcommand{\AFB}{\ensuremath{A_\mathrm{FB}}\xspace}
\newcommand{\EE}{\ensuremath{e^+e^-}\xspace}
\newcommand{\MM}{\ensuremath{\mu^+\mu^-}\xspace}
\newcommand{\TT}{\ensuremath{\tau^+\tau^-}\xspace}
\newcommand{\wangle}{\ensuremath{\sin^{2}\theta_{\mathrm{eff}}^\mathrm{lept}(M^2_\mathrm{Z})}\xspace}
\newcommand{\ttbar}{\ensuremath{{t\overline{t}}}\xspace}
\newcommand{\stat}{\ensuremath{\,\text{(stat.)}}\xspace}
\newcommand{\syst}{\ensuremath{\,\text{(syst.)}}\xspace}

\newcommand{\HGG}{\ensuremath{\mathrm{H}\to\gamma\gamma}}
\newcommand{\gev}{\GeV}
\newcommand{\GAMJET}{\ensuremath{\gamma + \mathrm{jet}}}
\newcommand{\PPTOJETS}{\ensuremath{\mathrm{pp}\to\mathrm{jets}}}
\newcommand{\PPTOGG}{\ensuremath{\mathrm{pp}\to\gamma\gamma}}
\newcommand{\PPTOGAMJET}{\ensuremath{\mathrm{pp}\to\gamma +
\mathrm{jet}
}}
\newcommand{\MH}{\ensuremath{\mathrm{M_{\mathrm{H}}}}}
\newcommand{\RNINE}{\ensuremath{\mathrm{R}_\mathrm{9}}}
\newcommand{\DR}{\ensuremath{\Delta\mathrm{R}}}


\newcommand{\PT}{\ensuremath{p_{\mathrm{T}}}\xspace}
\newcommand{\pt}{\ensuremath{p_{\mathrm{T}}}\xspace}
\newcommand{\ET}{\ensuremath{E_{\mathrm{T}}}\xspace}
\newcommand{\HT}{\ensuremath{H_{\mathrm{T}}}\xspace}
\newcommand{\et}{\ensuremath{E_{\mathrm{T}}}\xspace}
\newcommand{\Em}{\ensuremath{E\!\!\!/}\xspace}
\newcommand{\Pm}{\ensuremath{p\!\!\!/}\xspace}
\newcommand{\PTm}{\ensuremath{{p\!\!\!/}_{\mathrm{T}}}\xspace}
\newcommand{\ETm}{\ensuremath{E_{\mathrm{T}}^{\mathrm{miss}}}\xspace}
\newcommand{\MET}{\ensuremath{E_{\mathrm{T}}^{\mathrm{miss}}}\xspace}
\newcommand{\ETmiss}{\ensuremath{E_{\mathrm{T}}^{\mathrm{miss}}}\xspace}
\newcommand{\VEtmiss}{\ensuremath{{\vec E}_{\mathrm{T}}^{\mathrm{miss}}}\xspace}

%

\newcommand{\ga}{\ensuremath{\gtrsim}}
\newcommand{\la}{\ensuremath{\lesssim}}
\newcommand{\swsq}{\ensuremath{\sin^2\theta_W}\xspace}
\newcommand{\cwsq}{\ensuremath{\cos^2\theta_W}\xspace}
\newcommand{\tanb}{\ensuremath{\tan\beta}\xspace}
\newcommand{\tanbsq}{\ensuremath{\tan^{2}\beta}\xspace}
\newcommand{\sidb}{\ensuremath{\sin 2\beta}\xspace}
\newcommand{\alpS}{\ensuremath{\alpha_S}\xspace}
\newcommand{\alpt}{\ensuremath{\tilde{\alpha}}\xspace}

\newcommand{\QL}{\ensuremath{Q_L}\xspace}
\newcommand{\sQ}{\ensuremath{\tilde{Q}}\xspace}
\newcommand{\sQL}{\ensuremath{\tilde{Q}_L}\xspace}
\newcommand{\ULC}{\ensuremath{U_L^C}\xspace}
\newcommand{\sUC}{\ensuremath{\tilde{U}^C}\xspace}
\newcommand{\sULC}{\ensuremath{\tilde{U}_L^C}\xspace}
\newcommand{\DLC}{\ensuremath{D_L^C}\xspace}
\newcommand{\sDC}{\ensuremath{\tilde{D}^C}\xspace}
\newcommand{\sDLC}{\ensuremath{\tilde{D}_L^C}\xspace}
\newcommand{\LL}{\ensuremath{L_L}\xspace}
\newcommand{\sL}{\ensuremath{\tilde{L}}\xspace}
\newcommand{\sLL}{\ensuremath{\tilde{L}_L}\xspace}
\newcommand{\ELC}{\ensuremath{E_L^C}\xspace}
\newcommand{\sEC}{\ensuremath{\tilde{E}^C}\xspace}
\newcommand{\sELC}{\ensuremath{\tilde{E}_L^C}\xspace}
\newcommand{\sEL}{\ensuremath{\tilde{E}_L}\xspace}
\newcommand{\sER}{\ensuremath{\tilde{E}_R}\xspace}
\newcommand{\sFer}{\ensuremath{\tilde{f}}\xspace}
\newcommand{\sQua}{\ensuremath{\tilde{q}}\xspace}
\newcommand{\sUp}{\ensuremath{\tilde{u}}\xspace}
\newcommand{\suL}{\ensuremath{\tilde{u}_L}\xspace}
\newcommand{\suR}{\ensuremath{\tilde{u}_R}\xspace}
\newcommand{\sDw}{\ensuremath{\tilde{d}}\xspace}
\newcommand{\sdL}{\ensuremath{\tilde{d}_L}\xspace}
\newcommand{\sdR}{\ensuremath{\tilde{d}_R}\xspace}
\newcommand{\sTop}{\ensuremath{\tilde{t}}\xspace}
\newcommand{\stL}{\ensuremath{\tilde{t}_L}\xspace}
\newcommand{\stR}{\ensuremath{\tilde{t}_R}\xspace}
\newcommand{\stone}{\ensuremath{\tilde{t}_1}\xspace}
\newcommand{\sttwo}{\ensuremath{\tilde{t}_2}\xspace}
\newcommand{\sBot}{\ensuremath{\tilde{b}}\xspace}
\newcommand{\sbL}{\ensuremath{\tilde{b}_L}\xspace}
\newcommand{\sbR}{\ensuremath{\tilde{b}_R}\xspace}
\newcommand{\sbone}{\ensuremath{\tilde{b}_1}\xspace}
\newcommand{\sbtwo}{\ensuremath{\tilde{b}_2}\xspace}
\newcommand{\sLep}{\ensuremath{\tilde{l}}\xspace}
\newcommand{\sLepC}{\ensuremath{\tilde{l}^C}\xspace}
\newcommand{\sEl}{\ensuremath{\tilde{e}}\xspace}
\newcommand{\sElC}{\ensuremath{\tilde{e}^C}\xspace}
\newcommand{\seL}{\ensuremath{\tilde{e}_L}\xspace}
\newcommand{\seR}{\ensuremath{\tilde{e}_R}\xspace}
\newcommand{\snL}{\ensuremath{\tilde{\nu}_L}\xspace}
\newcommand{\sMu}{\ensuremath{\tilde{\mu}}\xspace}
\newcommand{\sNu}{\ensuremath{\tilde{\nu}}\xspace}
\newcommand{\sTau}{\ensuremath{\tilde{\tau}}\xspace}
\newcommand{\Glu}{\ensuremath{g}\xspace}
\newcommand{\sGlu}{\ensuremath{\tilde{g}}\xspace}
\newcommand{\Wpm}{\ensuremath{W^{\pm}}\xspace}
\newcommand{\sWpm}{\ensuremath{\tilde{W}^{\pm}}\xspace}
\newcommand{\Wz}{\ensuremath{W^{0}}\xspace}
\newcommand{\sWz}{\ensuremath{\tilde{W}^{0}}\xspace}
\newcommand{\sWino}{\ensuremath{\tilde{W}}\xspace}
\newcommand{\Bz}{\ensuremath{B^{0}}\xspace}
\newcommand{\sBz}{\ensuremath{\tilde{B}^{0}}\xspace}
\newcommand{\sBino}{\ensuremath{\tilde{B}}\xspace}
\newcommand{\Zz}{\ensuremath{Z^{0}}\xspace}
\newcommand{\sZino}{\ensuremath{\tilde{Z}^{0}}\xspace}
\newcommand{\sGam}{\ensuremath{\tilde{\gamma}}\xspace}
\newcommand{\chiz}{\ensuremath{\tilde{\chi}^{0}}\xspace}
\newcommand{\chip}{\ensuremath{\tilde{\chi}^{+}}\xspace}
\newcommand{\chim}{\ensuremath{\tilde{\chi}^{-}}\xspace}
\newcommand{\chipm}{\ensuremath{\tilde{\chi}^{\pm}}\xspace}
\newcommand{\Hone}{\ensuremath{H_{d}}\xspace}
\newcommand{\sHone}{\ensuremath{\tilde{H}_{d}}\xspace}
\newcommand{\Htwo}{\ensuremath{H_{u}}\xspace}
\newcommand{\sHtwo}{\ensuremath{\tilde{H}_{u}}\xspace}
\newcommand{\sHig}{\ensuremath{\tilde{H}}\xspace}
\newcommand{\sHa}{\ensuremath{\tilde{H}_{a}}\xspace}
\newcommand{\sHb}{\ensuremath{\tilde{H}_{b}}\xspace}
\newcommand{\sHpm}{\ensuremath{\tilde{H}^{\pm}}\xspace}
\newcommand{\hz}{\ensuremath{h^{0}}\xspace}
\newcommand{\Hz}{\ensuremath{H^{0}}\xspace}
\newcommand{\Az}{\ensuremath{A^{0}}\xspace}
\newcommand{\Hpm}{\ensuremath{H^{\pm}}\xspace}
\newcommand{\sGra}{\ensuremath{\tilde{G}}\xspace}
\newcommand{\mtil}{\ensuremath{\tilde{m}}\xspace}
\newcommand{\rpv}{\ensuremath{\rlap{\kern.2em/}R}\xspace}
\newcommand{\LLE}{\ensuremath{LL\bar{E}}\xspace}
\newcommand{\LQD}{\ensuremath{LQ\bar{D}}\xspace}
\newcommand{\UDD}{\ensuremath{\overline{UDD}}\xspace}
\newcommand{\Lam}{\ensuremath{\lambda}\xspace}
\newcommand{\Lamp}{\ensuremath{\lambda'}\xspace}
\newcommand{\Lampp}{\ensuremath{\lambda''}\xspace}
\newcommand{\spinbd}[2]{\ensuremath{\bar{#1}_{\dot{#2}}}\xspace}

\newcommand{\MD}{\ensuremath{{M_\mathrm{D}}}\xspace}
\newcommand{\Mpl}{\ensuremath{{M_\mathrm{Pl}}}\xspace}
\newcommand{\Rinv} {\ensuremath{{R}^{-1}}\xspace}

%
%
\hyphenation{en-viron-men-tal}

\cmsNoteHeader{09-007}
\title{CMS Data Processing
Workflows during an Extended Cosmic Ray Run}










\author{The CMS Collaboration}

\date{\today}

\abstract{The CMS Collaboration conducted a month-long data taking exercise, the Cosmic Run At Four Tesla, during October-November
2008, with the goal of commissioning the experiment for extended
operation.  With all installed detector systems participating, CMS
recorded 270 million cosmic ray events with the solenoid at
a magnetic field strength of 3.8~T. This paper describes the
data flow from the detector through the various online and offline computing
systems, as well as the workflows used for recording the data, for
aligning and calibrating the detector, and for analysis of the data.  }

\hypersetup{%
pdfauthor={R. Mankel, O. Gutsche et. al.},%
pdftitle={CMS Data Processing Workflows during an Extended Cosmic Ray Run},%
pdfsubject={CMS},%
pdfkeywords={CMS, physics, software, computing}}

\maketitle 


\section{Introduction}
\label{sec:intro}

The primary goal of the Compact Muon Solenoid (CMS)
experiment~\cite{cms} is to explore physics at the TeV energy
scale, exploiting the collisions delivered by the Large
Hadron Collider (LHC)~\cite{lhcEB}.
The central feature of the CMS apparatus is a
superconducting solenoid, of 6~m internal diameter. Within the field
volume are the silicon pixel and strip tracker, the crystal
electromagnetic calorimeter (ECAL) and the brass-scintillator hadronic
calorimeter (HCAL). Muons are measured in drift tube chambers (DT),
resistive plate chambers (RPC), and cathode strip chambers (CSC)
embedded in the steel return yoke. A detailed description of the experimental apparatus can be found
elsewhere~\cite{cms}.

A key element to the success of the experiment is the adequate design,
implementation and smooth operation of the data processing workflows
from the detector to the end user analysis. The month-long data taking
exercise known as the Cosmic Run At Four Tesla (CRAFT)~\cite{CRAFTGeneral} represented a major test for
these workflows. This paper describes the technical details of the
data flow from the detector to the final analysis. It explains the
data acquisition system and the various online and offline computing systems, and describes the software
and the workflows used in the data taking chain.

Section~\ref{sec:online} describes the online data taking environment
and Section~\ref{sec:hlt} the high-level trigger 
chain including the binary raw detector output content. The computing infrastructure
used to handle the recorded data is detailed in Section~\ref{sec:data}, and the
software and its special setup for reconstructing cosmic ray events in
Section~\ref{sec:reco}. This is followed by the description of the
data quality monitoring and the 
various validation steps performed during data taking in Section~\ref{sec:dqm}. The recorded data have been used to derive
alignment and calibration constants, which are described in Section~\ref{sec:alcareco}. The management and distribution of the constants via the CMS conditions
database system are addressed in Section~\ref{sec:conditions}, while the analysis
of the recorded cosmic ray muon data is described in Section~\ref{sec:analysis}.




\section{Online system}
\label{sec:online}
The CMS trigger and data acquisition (DAQ) system is designed to collect and 
analyse the detector information at the LHC bunch-crossing frequency of 40 MHz. The rate of events to be recorded for offline processing and 
analysis is about a few hundred Hz. The first-level trigger (L1) is designed to reduce the incoming data rate to a maximum of 
100 kHz, by processing fast trigger information coming from the calorimeters and the muon chambers, and selecting events with interesting signatures. 
The DAQ system must sustain a maximum input rate of 100 kHz, or a data flow of about 100 GB/s, coming from approximately 650 data sources from the different detector components. The DAQ system then reduces this rate by a factor of 1000 using a high-level trigger (HLT, Section~\ref{sec:hlt}), a
software filtering system running on a large processor farm.
\begin{figure}[htb!]
  \begin{center}
     \includegraphics[width=1.0\textwidth]{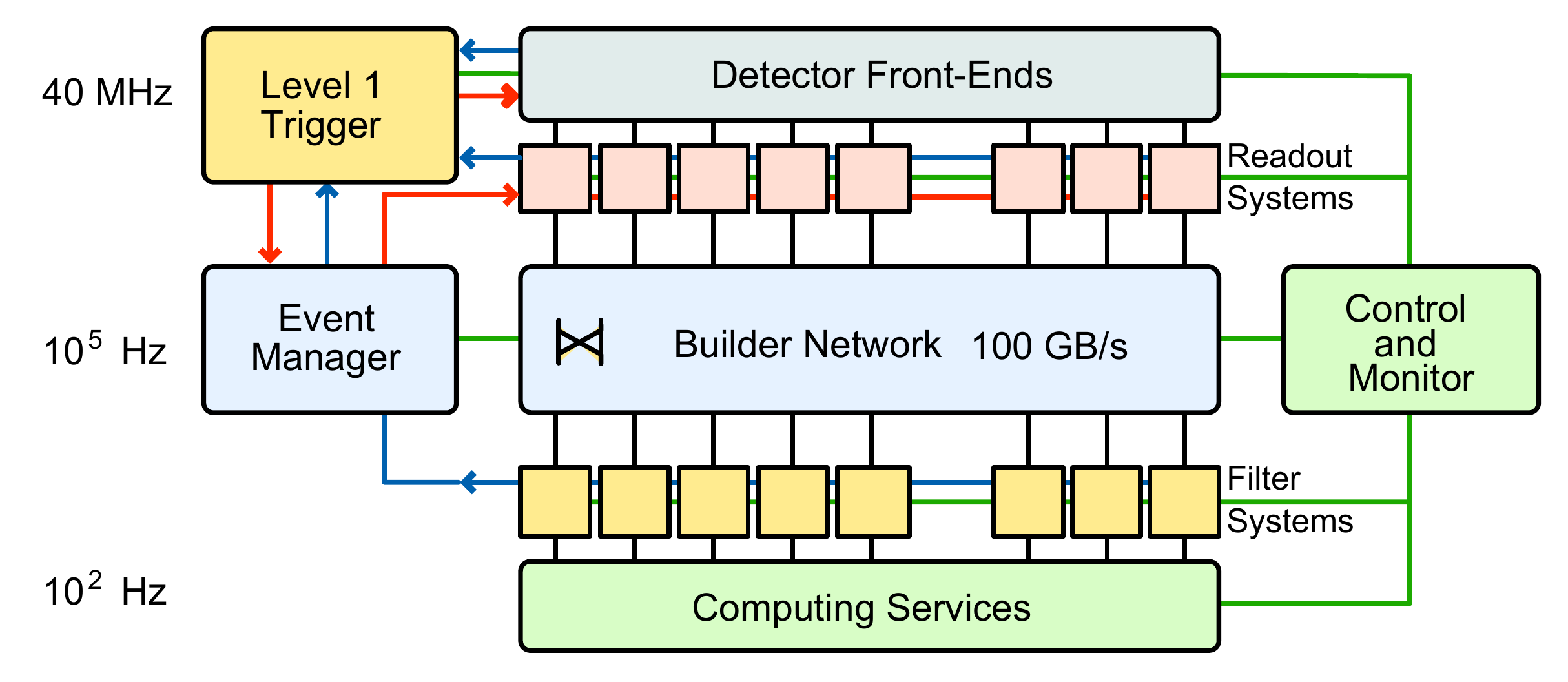}
   \caption[~Schematic view of the CMS DAQ]{Simplified schematic view of the CMS Data Acquisition System architecture. Shown are the key building blocks for a single slice of the DAQ system.}  
   \label{fig:daqarch}
  \end{center}
\end{figure} 

The architecture of the CMS DAQ system, described in detail elsewhere \cite{cms,daq}, is shown schematically in 
Fig.~\ref{fig:daqarch}. The sub-detector front-end systems store data continuously in 40 MHz-pipelined buffers. Synchronous L1 accept signals 
are distributed to the front-ends via the timing, trigger, and control system. When an accept signal is received, 
the corresponding data are extracted from the buffers in the front-ends and pushed into the DAQ system through the links in the readout systems.
The various readout fragments coming from different parts of the apparatus are subsequently assembled into complete events (event building) in two stages inside the high-performance builder network. Firstly, the front-end fragments 
are assembled into larger groups called {\it super fragments}, which are delivered to readout units organised in eight independent sets (DAQ slices). 
All super fragments belonging to the same event are given to the same slice which are fed with events in a round-robin fashion. 
In each slice, individual events are assigned to event buffers (builder units) by an event manager, and each builder unit assembles 
single events after having obtained all its super fragments from the readout units in the slice.
The builder unit hands over complete events to the filter systems running individual filter units upon request. The filter unit runs the HLT algorithms to 
select events to be accepted for storage, and eventually hands over accepted events to the computing services. In the end, storage managers, one for each slice, stream event data to disk and transfer complete data files to the CMS Tier-0 (Section~\ref{sec:data}).
More details about the HLT and data logging are discussed in the following section.


\section{High level trigger and data streams}
\label{sec:hlt}



The CMS high-level trigger algorithms are executed in a farm comprising 720 computing nodes, the event filter farm, executing the HLT reconstruction and selection algorithm sequence 
on individual events in parallel. The products of the HLT execution (e.g.\ reconstructed physics objects, tracks, etc.) 
can be added to the event before it is sent to storage, thus facilitating later debugging and analysis of the HLT 
performance. 

The HLT reconstruction uses the same  framework as the offline reconstruction~\cite{cmssw}. The HLT configuration (menu)
is delivered to the individual processes by the run control system ~\cite{rcms}. HLT configurations are managed by 
a configuration system designed around a relational \linebreak database abstraction of the individual components (reconstruction modules, 
filters, etc.) and their parameters~\cite{confdb}. An HLT menu consists of a set of {\it trigger paths}, each consisting of a sequence of 
reconstruction and selection modules. Each path is normally designed to select a specific physics signature (e.g.\ inclusive muon events).
Calibration and other conditions data are retrieved from the online database and distributed to the 
HLT processes by a hierarchy of cache servers, connecting to a redundant FroNTier server~\cite{FRONTIER} that provides uncomplicated web access to databases. It is used by the CMS software, on all tiers of the distributed-computing infrastructure of CMS, to retrieve calibration and alignment constants (Section~\ref{sec:data}).

Events accepted by the HLT are delivered to the storage manager system (SM) via the same switched network, used for event building.
The SM consists of 16 independent processes running 
on independent computers, and connected through a fibre-channel switch to eight disk arrays, for a total of 320 TB of disk space.
The SM is capable of an aggregate maximum throughput to disk of up to 2 GB/s while concurrently transferring complete data files 
to the Tier-0 at up to 800 MB/s.

For CRAFT, CMS operated 4 slices of the DAQ system using 275 computing nodes and 320 TB of available disk capacity for the SM system.

Routing of individual event data to files in the SM is driven by the definition of output {\it streams}, which group events selected 
by specific HLT paths. Several different streams are normally defined to group together events according to their offline usage (e.g.\ 
primary ``physics'' stream, ``express'' stream, calibration streams, etc.). The same path can feed multiple streams and hence, in general, 
individual streams can overlap. Within a stream, sets of paths selecting similar signatures (e.g.\ ``inclusive muons'', etc.) can be further 
grouped into primary datasets (PDs). A PD is defined as a subset of the stream consisting of the events satisfying a certain group of paths 
selected by that stream. The PD definition is subsequently used by the Tier-0 {\it repacking} step to split the contents of a stream into its component PDs (Section~\ref{sec:data}). Overlaps between streams affect the transfer bandwidth to the Tier-0 while overlaps between PDs primarily affect the disk and tape space consumption of the recorded data.
Both the stream and PD definition are intimately connected with the HLT menu, and hence the three are handled as a
unit. The same configuration management system is used to maintain and distribute them, and a single identification key is used by the HLT, the 
SM, and the Tier-0 to retrieve the relevant portion of the configuration from the database.




\subsection{Streams and primary datasets in CRAFT}

For CRAFT a primary physics stream and several alignment and
calibration streams had been defined (see
Section~\ref{sec:alcareco}) based on L1 and HLT trigger decisions. The physics stream was divided up into
three main physics primary datasets and several ``technical''
datasets, used for subdetector specific commissioning analyses.

The three physics primary datasets were:
\begin{itemize}
\item {\bf Cosmics}: all events satisfying at least one of the muon trigger 
paths. Those events are mainly used in cosmic ray data analyses. 
Its rate was around 300~Hz.

\item {\bf Calo}: all events satisfying all other L1 physics trigger paths. 
There was  no explicit exclusion of the cosmic ray events trigger path in the Calo 
dataset, but the overlap was minimal. The Calo dataset was mainly populated by
detector noise events and amounted to about 300~Hz.

\item {\bf MinimumBias}: all events selected either by a forward hadronic calorimeter
technical trigger or by one requiring a minimal activity in the pixel
tracker. The MinimumBias dataset contained also a fraction of random
triggers, useful for commissioning studies. Its rate was about 10~Hz.
\end{itemize}

The processing of the output streams and their splitting in primary datasets is described in the next section.





\section{Data Handling}
\label{sec:data}

The computing centre at CERN
hosts the Tier-0 of the distributed computing system of CMS
\cite{CompModel}. The CMS computing system relies on a distributed infrastructure of
Grid resources, services and toolkits, to cope with computing
requirements for storage, processing, and analysis.  It is based on
building blocks provided by the Worldwide LHC Computing Grid project (WLCG)
\cite{wlcg}. The distributed computing centres available to CMS around
the world are configured in a tiered architecture (as proposed in the
MONARC~\cite{monarc} working group), that behaves as a single coherent
system. The Tier-0 hosts the initial processing of data coming from
the detector and corresponds to about 20\% of all computing resources available to
CMS. The Tier-1 level takes care of subsequent processing and re-processing workflows
(Section~\ref{sec:t1}) and has approximately 40\% of the CMS computing resources available, while the Tier-2 level hosts
Monte Carlo ({\it MC}) simulation and analysis and uses the remaining $\sim 40$\% of all CMS computing resources.

All streams defined by the online system (Section~\ref{sec:online})
and the HLT (Section~\ref{sec:hlt}) are written in a binary data
format, referred to as {\em streamer files}. A transfer system copies the streamer files from the online systems at the detector site to the main CERN computing centre to be converted to a ROOT-based event data format
\cite{cmssw,root}, split into primary datasets and stored on tape. A first reconstruction is performed and its output is stored in separate datasets. The event content of the detector measurements is called the {\it RAW} data-tier and the output of the reconstruction pass is called the {\it RECO} data-tier.

The primary datasets are distributed amongst seven Tier-1
sites available to CMS for custodial storage and further processing. They are located in France ({\small T1\_FR\_IN2P3}), Germany ({\small T1\_DE\_FZK}), Italy ({\small T1\_IT\_CNAF}), Spain ({\small T1\_ES\_PIC}), Taiwan ({\small T1\_TW\_ASGC}), the United Kingdom ({\small T1\_UK\_RAL}), and the United States ({\small T1\_US\_FNAL}).

In a final step, datasets stored at the Tier-1 sites are served to Tier-2 centres, where the final analysis to extract physics results is performed.

\subsection{Tier-0 workflows}
\label{sec:t0}

The Tier-0 performs the conversion of the streamer files into the
ROOT-based event data format ({\it repacking}) and splits the
streams into primary datasets (Section
\ref{sec:hlt}). This is followed by the reconstruction of the primary datasets. In the case of CRAFT, only the three main physics primary datasets were reconstructed at the Tier-0.

A Python-based
processing system~\cite{t0} with an ORACLE database schema for state tracking ({\it
  T0AST}) is used to control a dedicated batch system queue (managed by LSF~\cite{lsf}) to split and process the data. The input and output files are handled by the
CERN Advanced STORage manager (CASTOR) mass storage system~\cite{castor}. Table~\ref{tab:t0} gives an overview
of the volumes of data produced from the central data-handling
perspective during CRAFT. CMS collected over 2 billion events including technical events for monitoring and calibrations purposes. During data taking, CMS recorded events without magnetic field and with the solenoid at
a magnetic field strength of 3.8~T.

\begin{table}[hbt!]
  \centering
  \caption[~Overview of data produced during CRAFT]{Overview of data produced during the CRAFT run, from the central data-handling perspective}
\ \\
  \label{tab:t0}
  \begin{tabular}{|l || r|}
    \hline
    Number of primary datasets produced & $11$ \rule{0cm}{14pt} \\ \hline
    Number of events recorded             &   2 $\times 10^9$  \rule{0cm}{14pt}\\ \hline
    Number of events in Cosmics primary dataset & 370 $\times 10^6$ \rule{0cm}{14pt}\\
    \hline
    Number of runs recorded & $239$ \rule{0cm}{14pt}\\ \hline
    Total data volume recorded and produced & $396$ TB \rule{0cm}{14pt}\\ \hline
    Total data volume recorded and produced  in Cosmics primary dataset & $133$ TB \rule{0cm}{14pt} \\
    \hline
  \end{tabular}
\end{table}

In subsequent steps, the output of the reconstruction is used to derive specialised alignment and calibration datasets (Section~\ref{sec:alcareco}) and data quality
information is extracted and uploaded to a web server (see
Section~\ref{sec:dqm}).

For these two processing steps, the system normally used for MC production and central processing at the Tier-1 sites was
used as a temporary solution (\cite{prodagent}, Section~\ref{sec:t1}). The processing of newly recorded data was triggered periodically during data taking to produce alignment and calibration datasets and upload quality information to a web server. These functionalities are now integrated in the Python-based Tier-0 processing system.

\begin{figure}[htb!]
  \begin{center}
     \includegraphics[width=0.8\textwidth]{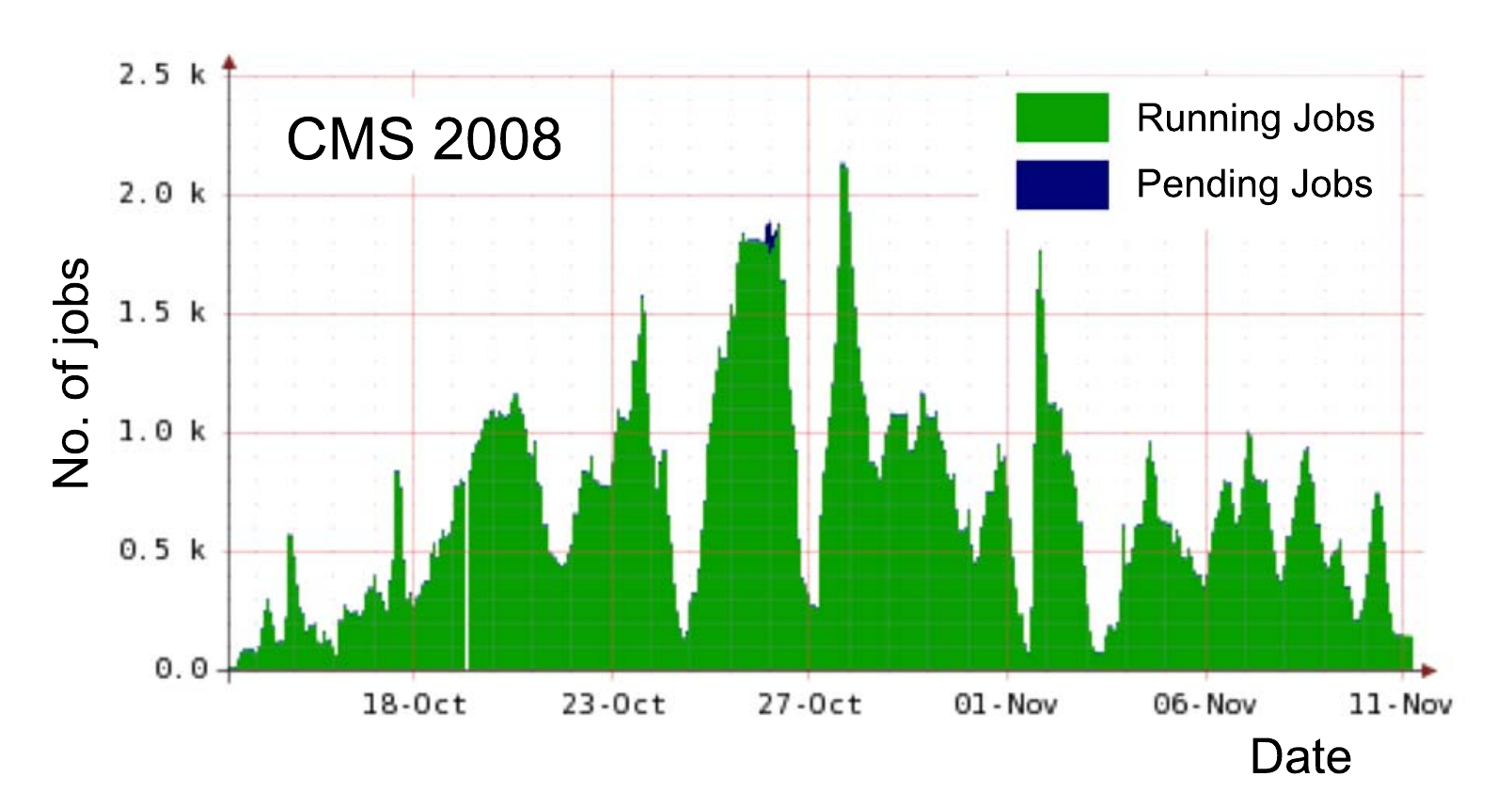}
  \caption[~Tier-0 LSF queue utilization]{Utilization of the Tier-0 LSF batch
    queue during CRAFT. The maximum of 2250 batch slots were never fully utilized and all jobs started promptly after submission. (Taken from monitoring sources).}
  \label{fig:t0_queue}
  \end{center}
\end{figure}

The LSF queue at the Tier-0 was dimensioned for early data taking in
Sept. 2008 and allowed for a maximum of 2250 jobs running in
parallel. As shown in Fig.~\ref{fig:t0_queue}, this capacity was never used completely  owing to the short reconstruction time of the low-occupancy cosmic ray events. The average reconstruction time in CRAFT was about 0.75 seconds per event. This should be compared to the reconstruction time of proton-proton collision events, which is estimated to be about 5 seconds per event. This estimate was derived from events producing top quark pairs, which do not represent the bulk of the expected events but resemble many of the physics processes of interest for analysis.

\begin{figure}[htb!]
  \begin{center}
     \includegraphics[width=0.8\textwidth]{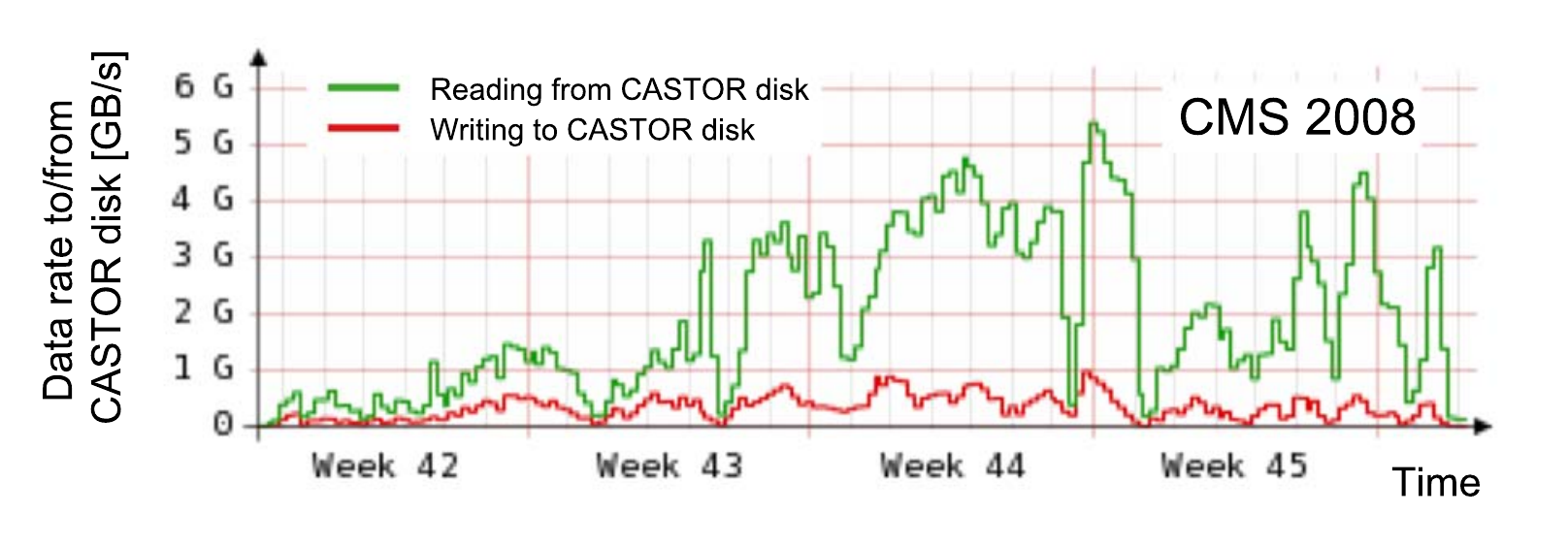}
  \caption[~Tier-0 MSS network utilization]{In-bound and
out-bound network data rates to the Tier-0 CASTOR disk pool at CERN. The
reading rate peaked at 5 GB/s while the writing rate was always below
1 GB/s. (Taken from monitoring sources).}
  \label{fig:t0_network}
  \end{center}
\end{figure}

The in-bound and out-bound network data rates to the Tier-0 CASTOR disk pool are shown in
Fig.~\ref{fig:t0_network}. The reading rate peaked at 5 GB/s
while the writing rate was always below 1 GB/s. The available network bandwidth was able to support all Tier-0 cosmic ray data taking workflows.

Overall, the Tier-0 infrastructure performed stably and reliably, and
was sufficiently provisioned for cosmic ray data taking.

\subsection{Data storage and transfers}
\label{sec:transfers}

The CMS computing model foresees at least two copies of all data on independent storage media, for example on tape at two different sites. To fulfill this requirement, all output datasets in ROOT format are stored on tape at the Tier-0. This copy, called the {\it archival} copy of the data, is only stored for backup purposes and is not accessible for processing workflows. A further copy is distributed amongst the Tier-1 sites for custodial storage on tape. This copy is called the {\it primary} copy and access is provided for further processing on the Tier-1 level (Section~\ref{sec:t1}) and analysis on the Tier-2 level. As a safety measure during CRAFT, all streamer files were stored temporarily on tape as well.

A data-transfer
management system named PhEDEx (Physics Experiment Data Export)
\cite{phedex} is used to handle the movement of data
between these computing centres. Deployed at all CMS sites, PhEDEx automates many
low-level tasks, such as large-scale
data replication and tape migration, and guarantees consistency of the dataset copies. PhEDEx
uses standard WLCG transfer tools
such as FTS~\cite{fts} and SRM~\cite{srm}, which interface with the mass storage
systems at Tier-1 and Tier-2 centres. PhEDEx provides site managers and users with a centralised
system for making data movement requests and provides status and
overview information.

During CRAFT, the recorded and processed primary datasets were
distributed amongst the Tier-1 sites according to available free tape
space taking into account processing capacity and reliability of the
Tier-1 sites. For the Cosmics primary dataset, the average size per event for the RAW data tier was 105 kB/event and for the RECO data tier 125 kB/event.

\begin{figure}[htb!]
  \begin{center}
     \includegraphics[width=0.8\textwidth]{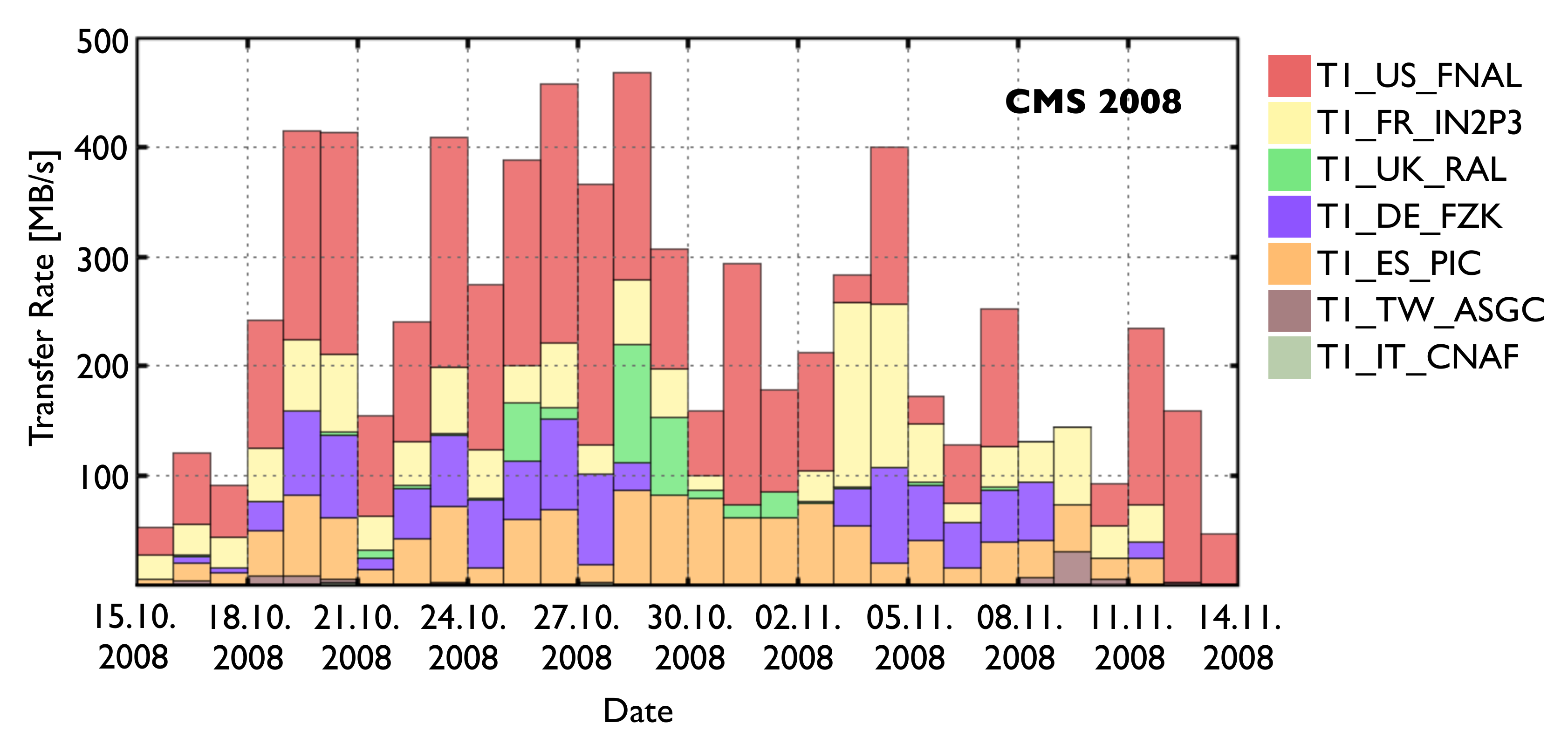}
  \caption[~Transfer rates from Tier-0 to Tier-1 centres]{Transfer
    rates from Tier-0 to Tier-1 centres over the duration of
    CRAFT. The average was about 240 MB/s. (Taken from monitoring sources).}
  \label{fig:transfer_rate}
  \end{center}
\end{figure}

Figure~\ref{fig:transfer_rate} shows the transfer rate during CRAFT from the Tier-0 to the Tier-1
sites. The transfers averaged $240$ MB/s with rates exceeding $400$
MB/s on several occasions.


\begin{figure}[htb!]
  \begin{center}
     \includegraphics[width=0.8\textwidth]{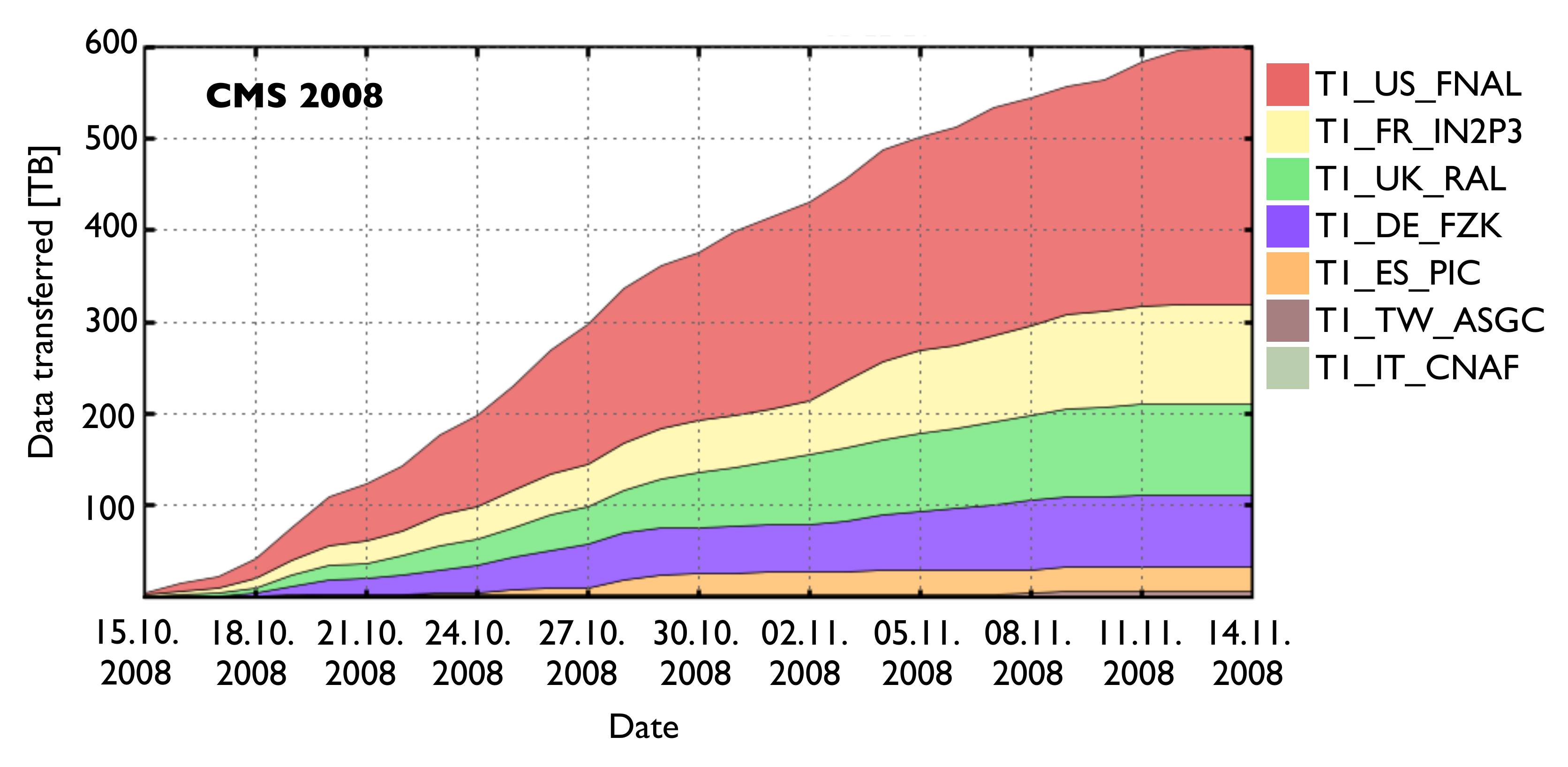}
  \caption[~Cumulative transfer volume from Tier-0 to Tier-1
  centres]{Cumulative transfer
    volume from Tier-0 to Tier-1 centres over the duration of
    CRAFT. (Taken from monitoring sources).}
  \label{fig:transfer_volume}
  \end{center}
\end{figure}

During CRAFT, a total of 600 TB was transferred out of CERN to the Tier-1
sites. Figure~\ref{fig:transfer_volume}
shows the cumulative transfer volume per Tier-1 site.


Overall, the
transfer system performed very well and transferred all CRAFT data
reliably to the Tier-1 sites. There was one very large 200~GB file failing
transfer to the US Tier-1 site. Being about $20$~times larger than the average file size in
the dataset, a timeout occurred during each transfer attempt
and caused a short delay in analysis of the
dataset. Later analysis on the file was performed and
safeguards have since been put into place to prevent similar failures
from occurring.

\subsection{Tier-1 processing}
\label{sec:t1}

The central processing at the Tier-1 sites was performed using the
current MC production system~\cite{prodagent}. The system is written in Python~\cite{Python} and uses a MySQL database~\cite{Mysql} to schedule jobs by interfacing with
different Grid middlewares~\cite{egee,nordugrid,osg}. However, compared to the Tier-0
system, it does not track the state of every processing step
in detail. It was optimised for production of Monte Carlo samples, for
which 100\% accountability
is not a primary concern because more
events can be easily generated in the case of failures or
infrastructure problems.

The requirements for the Tier-1 processing workflows are in fact very different compared to MC production workflows. During processing on the Tier-1 level, an input RAW or RECO dataset is
processed to either prepare a new reconstruction pass with updated
software and/or conditions and alignment constants ({\it
  re-reconstruction}) or to extract events of interest from the total
dataset to reduce the amount of data to be analysed ({\it
  skimming}). Accounting
of the processing of every single input event is of highest priority
as all events have to be accounted for in order to correctly calculate the luminosity for the processed LHC collisions.

In the near future, the used Monte Carlo production system will be completely redesigned based on the state tracking technology of the Tier-0 processing system. Guaranteeing 100\% accountability, the new system is planned to be put into operation in spring 2010.


In CRAFT, the physics primary datasets were skimmed to reduce the amount of data for physics analyses to
process, still keeping full physics content. Some skims combined the RAW and RECO event content in the same dataset to simplify debugging of the software and low-level analysis of the events. In addition, many skims
were used for re-reconstruction to validate new algorithms
or new derived alignment and calibrations conditions thus reducing the
number of full reprocessings.

The list of skims exercised during CRAFT is shown in Table~\ref{tab:WF_skims} by listing the parent primary dataset and acceptance of the skim event selection as well as the output event content of each skim.
\begin{table}
\centering
\caption{List of skims exercised during CRAFT, showing for each skim the parent primary dataset, acceptance of the skim event selection and output event content combined from different data tiers. The {\it RAW} data-tier consists of all detector measurements while the {\it RECO} data-tier contains all reconstructed properties of an event.
\label{tab:WF_skims}}
\begin{tabular}{|l||c|c|c|}
\hline
Skim name             & Prim. Dataset & Acceptance & Event Content \\
\hline
SuperPointing         &  Cosmics  & 0.27\%  & RAW-RECO  \\
TrackerPointing       &  Cosmics  & 4.50\%  & RAW-RECO  \\
MultiMuon         &  Cosmics  & 0.44\%  & RECO  \\
PtMinSelector         &  Cosmics  & 0.68\%  & RECO \\
\hline
CSC\_Skim         &  Cosmics  & 3.84\%  & RAW-RECO \\
CSC\_Skim\_BFieldStudies &  Cosmics  & 0.04\%  & RAW-RECO \\
\hline
HCALHighEnergy        &  Cosmics/Calo/MinBias  & 0.09\% (on Cosmics) & RAW-RECO \\
ECALSkim              &  Cosmics/Calo  & 0.40\% (on Cosmics) & RECO \\
\hline

\end{tabular}
\end{table}

The {\it SuperPointing} and {\it TrackerPointing} skims preferentially selected events containing muons whose propagation
loosely traversed the Pixel and Strip tracker regions, respectively. The {\it MultiMuon} skim
contained events with more than four reconstructed muons for muon shower analyses, while the {\it PtMinSelector} skim retained only events with high-energy muons ($p_T>50$~GeV/c).

The {\it CSC\_Skim} skim selected events with activity in the CSC endcap muon detectors. In addition, good segments from CSC hits were selected in the
{\it CSC\_Skim\_BFieldStudies} skim, which were used in a measurement of
the magnetic field in the endcap yoke regions.

The {\it HCAL} skim selected events with high energy deposits in the HCAL and the {\it ECAL} skim was selecting events with high energy deposits in the ECAL due to either showering muons or noise activity.

Two re-reconstruction passes of all data taken in CRAFT
were made after the data taking period ended. The
first pass was started in November 2008 and finished in January 2009. It suffered significantly from infrastructure and technical problems. After improvements, a second pass was performed from the
12$^{\rm th}$ to the 25$^{\rm th}$ of February 2009. Both
re-reconstruction passes produced the associated
alignment and calibration datasets, as was done for the Tier-0 processing.

The major issue observed during the re-reconstruction was the lack of complete
accountability of the reprocessing system. This complicated the task
of identifying job failures, which had to be done manually. The net
result was that a very small fraction (about a few percent) of the
final processing was incomplete.

This will be addressed in the future by the new Tier-1 processing system, developed for LHC collision running.



\section{Reconstruction Software}
\label{sec:reco}

The main goals of event reconstruction in the context of CRAFT were to
provide a reliable cosmic ray event reconstruction in order to support
detector performance, calibration, and alignment studies, as well as 
testing as much as possible the software components to be used in
proton-proton collision events with the LHC.

In order to accomplish these objectives, a dedicated configuration of
the reconstruction software tools was prepared and optimised for
cosmic ray muon events. While some part of the reconstruction code developed for
proton-proton collisions could be basically re-used, many key
 elements needed additional code development to support the
reconstruction of cosmic ray muon events. The code developed for
proton-proton collision events is nevertheless used and tested wherever
possible. In several cases, two or more reconstruction algorithms performing similar tasks have been run
in parallel to gain experience (Table~\ref{tab:components}).
\begin{table}
\begin{center}
\caption{Summary of the different reconstruction algorithms used
during CRAFT. The central column indicates the main algorithm executed 
as input to subsequent steps. The right column
shows the alternative algorithms performed in parallel. See Section~\ref{sec:algo} for details.}
\label{tab:components}
\ \\
\begin{tabular}{|l|l|l|}
\hline
\bf{Component} & \bf{Default Code/Configuration} & \bf{Alternative versions} \\ 
\hline
\hline
Tracker local & standard & none \\ 
\hline
ECAL & pulse fit based & standard weight based \\ 
\hline
HCAL & threshold for cosmic ray events & none \\ 
\hline
DT  & standard & no-drift (coarse hits) , $t_0$ fitting  \\ 
\hline
CSC & standard & none \\ 
\hline
RPC  & standard & none  \\ 
\hline
Tracking & dedicated seeding & Road Search (cosmic version) \\ 
   &  and navigation & Cosmic Track Finder (no pattern reco)\\
   &  & track splitting (up/down)\\
\hline
Muon  & dedicated cosmic & barrel/endcap only \\ 
   &  muon reconstruction & different DT local reco\\
   &  & LHC vs.\ Cosmic navigation and fit\\
   &  & single-leg vs.\ two-leg\\
\hline
Jet and MET  & standard & none \\
\hline
Electron/Photon  & subset of standard & none \\
\hline
B tagging  & not run & none \\
\hline
Particle Flow and & not run & none \\
Tau reconstruction & &  \\
\hline

\end{tabular}
\end{center}

\end{table}

In the following section we briefly describe the major changes in
reconstruction code during the CRAFT data taking period. 
\subsection{Dedicated reconstruction algorithms for cosmic ray events}
\label{sec:algo}

\subsubsection*{Local reconstruction in the electromagnetic calorimeter}
In parallel with the standard ECAL local reconstruction, an additional
algorithm has been used to provide a better measurement of the energy deposited
by a cosmic ray muon \cite{ecalrecopaper}. The standard algorithm is optimised for particles
reaching the ECAL crystals within a very narrow time window, as
appropriate for LHC collisions, while in case of cosmic ray muons the
actual time of arrival is spread over the 25~ns of the bunch clock
interval. The modified algorithm performs a fit of the pulse shape,
sampled tenfold by the ECAL hardware, and thus provides a precise
measurement of both the energy and the arrival time of the cosmic ray
muon.  While particles from LHC interactions release energy in only a
few crystals, cosmic ray muons cross the ECAL surface at a wide range of
incidence angles, which can spread the energy deposit over a sizable
number of crystals. A dedicated version of the clustering
algorithm has been used to collect this energy most efficiently.
\subsubsection*{Track reconstruction within the Tracker}
The track reconstruction for CRAFT is largely based on the methods
already employed for the Magnet Test and Cosmic Challenge (MTCC) in
2006 \cite{bib:mtcctk,bib:mtcc}, and the Tracker Integration Facility (TIF)
sector test \cite{bib:tif} in Spring 2007. The main differences
compared to standard tracking for collisions are in the \emph{seeding}
and in the \emph{navigation} steps. Seeding combines hits from several
neighbouring layers to generate the starting point for the track
pattern recognition and is mainly based on the pixel system in case of
LHC collision events; this has been modified to be able to reconstruct 
those trajectories not crossing the very small pixel volume. The modified
seeding mainly uses those layers of the silicon strip tracker that 
provide a three dimensional position measurement.  Navigation, on the
other hand, provides the set of the paths the particle can possibly
have taken between one tracking layer and another; these sets differ
considerably between cosmic ray muons and particles originating from the
central interaction point. 

For diagnostic purposes, the concept of top-bottom track splitting has
been introduced, in which the incoming and outgoing part of the cosmic
ray muon trajectory, with respect to the point of closest approach to the
beam line, are reconstructed as separate tracks. Comparison of the
parameters of these two {\em legs} serves as a powerful instrument for
alignment and tracking performance studies~\cite{bib:alignmentnote}.

\subsubsection*{Reconstruction of muons}
This section describes the reconstruction of muon trajectories in
the CMS muon system. When possible,
the muon system part of the trajectory is combined with a track
reconstructed within the tracker, resulting in a {\em global muon
track}.


Several different flavours of muon reconstruction are used combining
different configurations of the various components. As described in Ref.~\cite{bib:muoncraftreco} it is possible to perform dedicated cosmic
ray muon reconstruction as a single track (referred to as {\em single leg}
mode), or split the trajectory into an incoming and outgoing track
({\em two legs}). Alternatively, the standard reconstruction for
collision events can be used, in which optionally the reconstruction
can be restricted to either the barrel or the endcap regions.

Muon reconstruction also depends on the
local reconstruction of the DT, RPC, and CSC sub-detectors, for which
different variants are available. For reconstruction of the DT track
segments, two options are addressing the fact that cosmic ray 
muons, contrary to those from collisions, arrive at an arbitrary time
not correlated to the bunch clock. The {\em $t_0$-corrected}
segment reconstruction treats the arrival time $t_0$ as an additional
free parameter in the fit, while the {\em no-drift} variant does not
use drift time information at all, resulting in relatively coarse
point resolution.

The final \emph{reconstructed muon} object combines all the
information available for a reconstructed muon, including standalone
muon reconstruction (using only DT, RPC, CSC), global muon
reconstruction (matching the track from standalone muon reconstruction
with the silicon tracker track), and tracker-based reconstruction
(matching the tracker tracks with muon segments).

\subsection{Deployment of software updates during prompt reconstruction}
In view of the large-scale commissioning nature of CRAFT, prompt
reconstruction is one of the workflows for which a low latency of code
corrections is important. Fast deployment of bug fixes
may generally become necessary when new major releases have been
deployed for the first time, when running conditions change
drastically, or in the early stage of a data taking period. Most of
the problems encountered during CRAFT were related to unexpected
detector inputs and lack of corresponding protections in the code to
handle them properly.

The procedure for deploying an update was handled by several different shift roles with different responsibilities. The prompt reconstruction operator reported job failures. These failures were investigated by the offline run manager who identified the code to be fixed and contacted the appropriate experts. A minimal bug
fix was usually provided within a few hours; if no correction could be
achieved on this time scale, the corresponding reconstruction feature
was disabled to restore Tier-0 operations. This update was provided
either as a new configuration or as a new software release. 

The CRAFT experience has been a driving element for CMS to introduce a
patch-release system to reduce the time required for deployment of new
code. It is also seen as an advantage to be able to pause the Tier-0
processing for about 24~hours if necessary to take actions.  These
two points, combined with the fast feedback that an additional express
stream processing will provide, should
allow minimising the Tier-0 inefficiency due to reconstruction
software failures in future data taking.

\subsection{Evolution of Tier-1 reprocessing}
The two reprocessing cycles executed at Tier-1 centres allowed the reconstruction to be rerun with updated detector conditions (see
Section~\ref{sec:AlCaReprocessing} for details) and to improve the reconstruction code. In addition, these
reprocessing steps were used to tailor slightly the content of the reconstruction output files
according to requests from the commissioning and
alignment groups. The biggest changes concerned the muon
reconstruction and the track reconstruction in the tracker.


\section{Data Quality Monitoring and Prompt Feedback}
\label{sec:dqm}
Data quality monitoring is critically important for ensuring a good
detector and operation efficiency, and for the reliable certification
of the recorded data for physics analyses. The CMS-wide DQM system
comprises:

\begin{itemize}
  \item tools for creating, filling, transporting, and archiving of histograms
    and scalar monitor elements, with standardised algorithms for performing
    automated quality and validity tests on distributions;
  \item online monitoring systems for the detector, trigger, and
    DAQ hardware status and data throughput;
  \item offline monitoring systems for reconstruction
    and for validating calibration results, software releases, and simulated
    data;
  \item visualisation of the monitoring results;
  \item certification of datasets for physics analyses.
\end{itemize}

\begin{figure}[!b]
\begin{center}
\includegraphics[angle=-90,width=.9\textwidth]{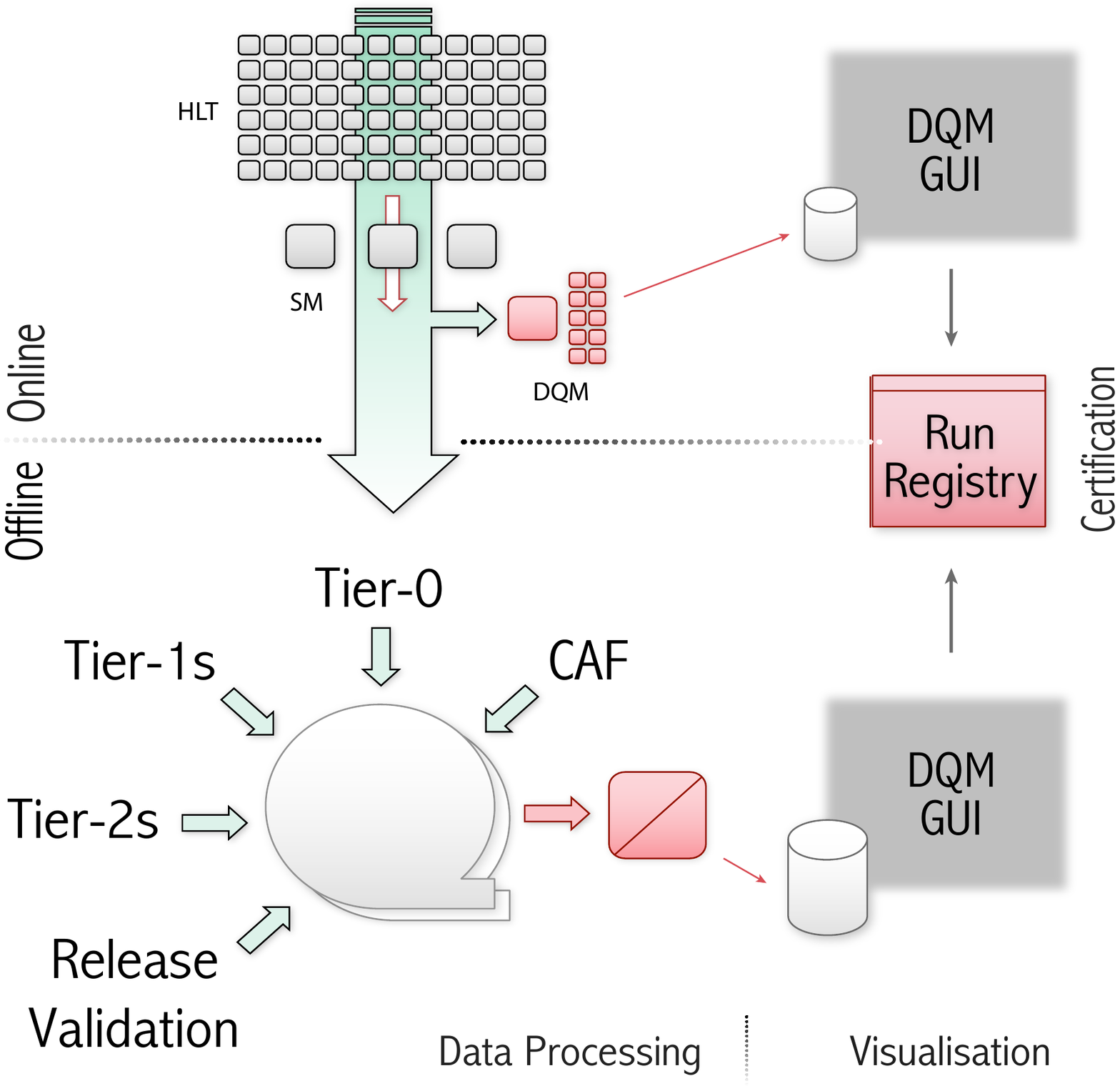}
\end{center}
\caption{\label{fig:overview}Sketch of the DQM system, consisting of branches
for online and offline monitoring.}
\end{figure}



The main features of the DQM system, as operated during the CRAFT data
taking period, are shown in Fig.~\ref{fig:overview}. A detailed
experience report can be found elsewhere~\cite{dqm_chep09}. DQM for
data taking is performed in two different stages, with very
small latency during the data taking (online) and after prompt
reconstruction of the events (offline).

\subsection{Online monitoring}

The online DQM system consists of a number of consumer applications,
labelled as DQM in Fig.~\ref{fig:overview}, usually one per
subsystem, which receive event data through a storage manager event
server and fill histograms at an event rate of 10--15~Hz.

In addition, a small number of histograms is filled in the
HLT filter units, which process events at up
to 100~kHz. These histograms are shipped out to DQM consumer
applications periodically. Identical histograms are summed up
across different filter units in the storage manager.

All the histogram data, including alarm states based on quality
test results, are made available to a central DQM graphical user interface server (GUI)  for visualisation
in real time~\cite{dqm_gui_09}, and are stored in a
ROOT file periodically during the run.  At the end of the run the final
archived results are uploaded to a large disk pool.
Eventually, the files are merged and backed up to tape.

All DQM data processing components and the event display
start and stop automatically under centralised CMS run control~\cite{runcontrol}.
The web servers for the DQM GUI~\cite{dqm_gui_09} and web-based conditions
monitoring (WBM)~\cite{wbm} are long-lived server systems which are
independent of the run control.


\subsection{Offline monitoring}

The offline DQM system accumulates monitoring data from several
workflows in CMS, namely Tier-0 prompt reconstruction, re-reconstruction at
the Tier-1s and the validation of the alignment and calibration results,
the software releases, and all the simulated data.

CMS has standardised the monitoring of the event data processing
into a two-step workflow:

\begin{enumerate}
\item The histogram monitor elements are created and filled with
CMS event data information. The histograms are stored along with the processed
events into the normal output event data files. When the CMS data processing
systems merge output files, the histograms are automatically summed
to form the first partial result.

\item At the end of the data processing the histograms are extracted
from the job output data files and summed together across entire runs
to yield full event statistics. The final histograms are then used to
calculate efficiencies and are checked for quality, by making comparisons
with reference distributions.
The histograms, certification results, and quality test results
are saved into a ROOT file, which is then uploaded to a central DQM GUI
web server. In the web server, the files are merged and backed up to
tape; recent data are kept cached on disk for several months.
\end{enumerate}

Online and offline DQM GUI web servers provide a common interface,
and are linked together as one entity, giving the entire world-wide
collaboration access to inspection and analysis of all DQM data
at one central location.

\subsection{Data certification}\label{certification}

CMS uses a database with a front-end web application, the run registry,
as the central workflow tracking and bookkeeping tool
to manage the creation of the final physics dataset certification result.
The run registry is both a user interface managing the workflow and a
persistent storage of the information.

The work to evaluate the detector and physics object data quality is organised
in shifts. The shift persons follow instructions specifically tailored to catch
problems. The observations are entered in the run registry database where they
are available to detector and physics object groups, as well as the whole
collaboration, for inspection and confirmation.
Final certification results are produced at regular sign-off meetings,
typically once per week, before they are delivered to the experiment
by storage in the data bookkeeping system (DBS)~\cite{DBS}.
The information in DBS is associated with the primary datasets
and is input to the creation of analysis datasets by analysis groups.

Online shifts take place continuously during detector operation at the CMS detector site.
Offline DQM shifts are carried out at daytime at the CMS centre~%
\cite{cms_centres_09} on the main CERN site.  The shift activities are
also supported by regular remote shifts, two shifts per day at Fermilab and one
shift per day at DESY, at the local CMS
centres~\cite{collaboration_at_distance_09}.

\subsection{Prompt data analysis and feedback}

The central DQM shift activity is complemented by the prompt feedback
groups, one for each subsystem, which consist of subsystem experts
located at the CMS centre. These groups analyse the prompt
reconstruction output and integrate the data quality
information in a timely way. The CMS CERN Analysis Facility (CAF)~\cite{caf}, providing large CPU power and
fast access to the data stored on a local CASTOR disk pool~\cite{castor},
was heavily used for such analyses.

\begin{figure}
\centering
\begin{minipage}[t]{0.45\textwidth}
\includegraphics[width=0.99\textwidth]{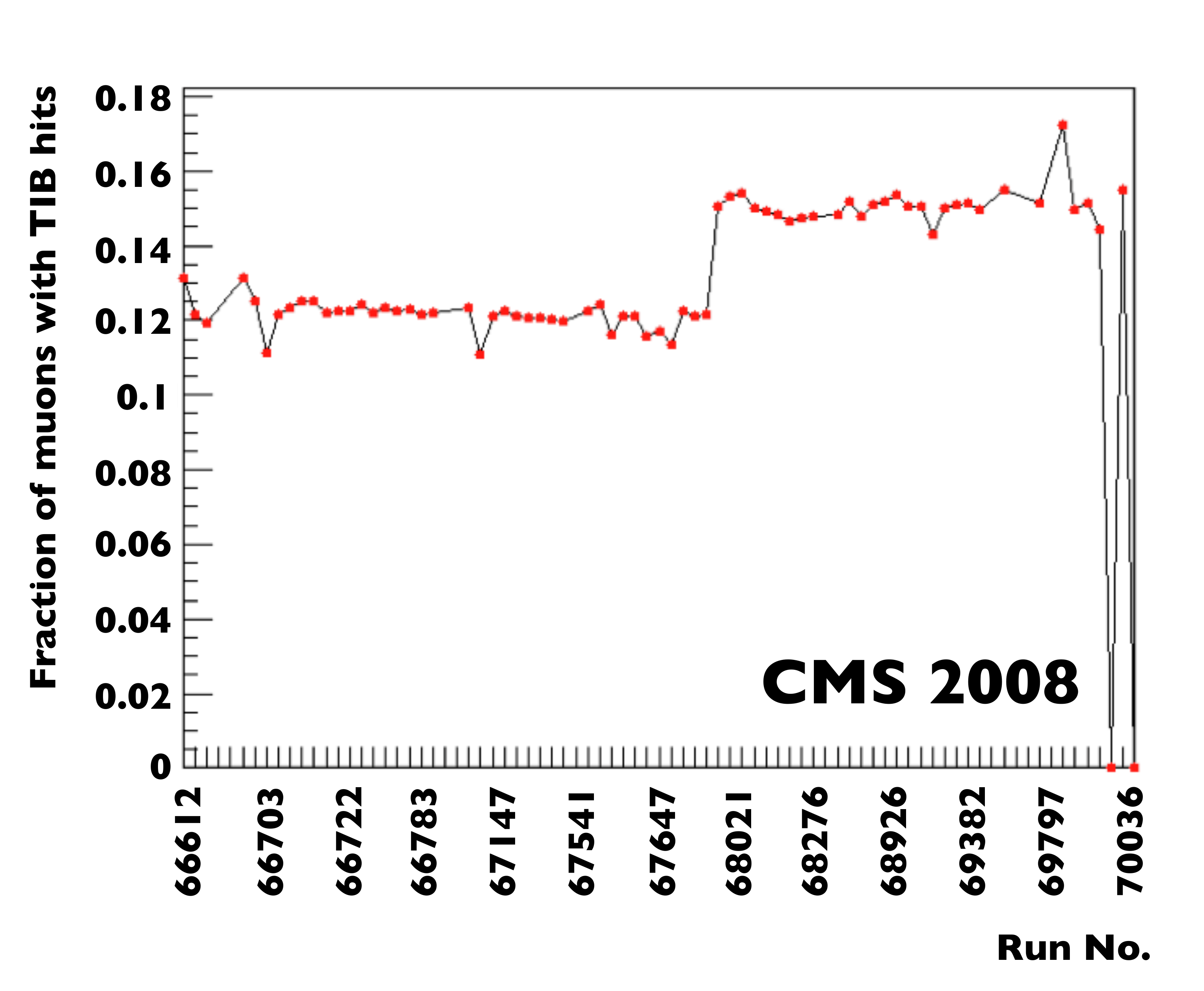}
\caption{Evolution of the relative number of hits on muon tracks in the tracker
inner barrel detector. The step in the distribution is due to
improved alignment parameter errors applied to the later data.
 \label{fig:histodqmtrack}}
\end{minipage}
\hspace{1em}
\begin{minipage}[t]{0.45\textwidth}
\centering
\includegraphics[width=0.99\textwidth]{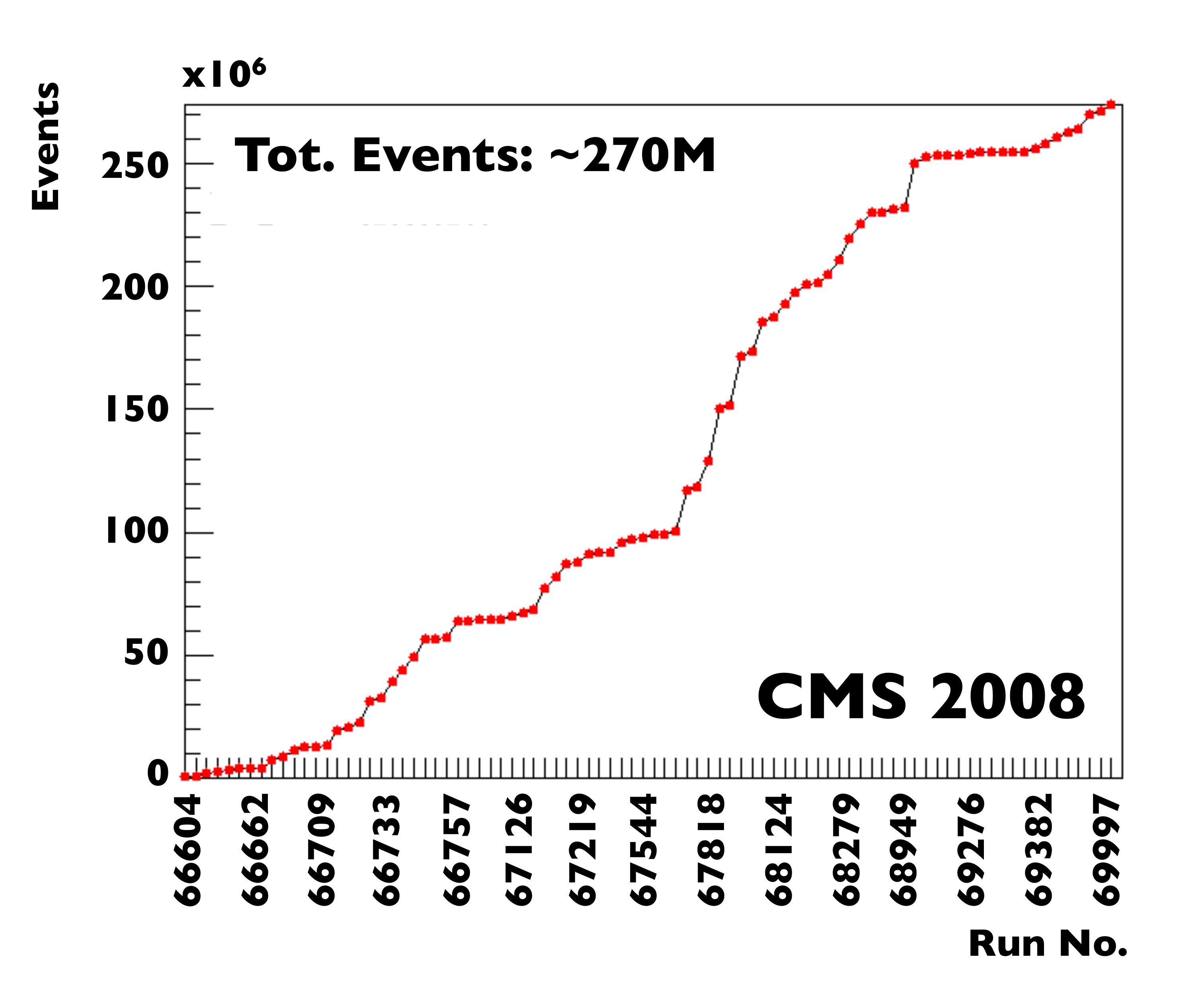}
\caption{Integrated statistics vs.\ run collected during CRAFT in the Cosmics dataset for runs with
good quality flags from the drift tubes and the silicon strip tracker. Only runs
with magnetic field of 3.8~T have been considered.
 \label{fig:eventsvsrun}}
\end{minipage}
\end{figure}

The run-by-run DQM results were used to monitor the
time evolution of the detector behaviour.
Any observed change was carefully checked and tracked.
As an example, Fig.~\ref{fig:histodqmtrack} shows the evolution of the
relative number of hits on tracks in the tracker inner barrel detector
as a function of the run number. The step in the distribution is due
to improved alignment parameter errors applied to the later data.
During reprocessing, the improved parameters were applied to all data,
thus removing the step.

Table~\ref{tab:allevents} shows the number of good events in the
Cosmics primary dataset, based on the quality assignment described
above. The breakdown for each subsystem is given when operating nominally and passing the offline selection criteria. Most of the subsystems declared individually more than 85\% of the
recorded data in the Cosmics primary datasets as good. Having declared a detector component as good does not entail having detected a cosmic ray muon within it's fiducial volume. Figure~\ref{fig:eventsvsrun} shows the accumulated number of cosmic ray triggered events as a function of run number with the magnet at its operating central field of 3.8~T, where the minimal configuration of the silicon strip tracker and the DT muon system delivering data certified for further offline analysis was required. It was not required to keep the other systems in the configuration. A total of 270 million such events were collected.


\begin{table}
\centering
\caption{Cosmic ray triggered events collected during CRAFT in the Cosmics primary dataset in
periods when the magnetic field was at the nominal value of 3.8~T with the listed detector system (or combination of systems) operating nominally and passing offline quality criteria. The minimum configuration required for data taking was that at least the DT barrel muon chambers and the strip tracker passed the quality criteria. The other subdetectors were allowed to go out of data taking for tests.
\label{tab:allevents}}
\ \\
\begin{tabular}{|l|c|}
\hline
Quality flag & Events (millions) \\
\hline
\hline
 (none) & 315 \\
\hline
Trigger & 240  \\
Pixel Tracker & 290 \\
Strip Tracker & 270 \\
ECAL & 230 \\
HCAL & 290 \\
RPC & 270 \\
CSC & 275 \\
DT & 310 \\
DT+Strip & 270 \\
All & 130\\
\hline
\end{tabular}
\end{table}

\section{Alignment and Calibration}
\label{sec:alcareco}
This section describes the workflows used to compute and improve
the alignment and calibration constants. While some calibrations are
already performed online at the CMS detector site, this
section will focus on the workflows performed at the Tier-0 site and
at the CAF.

\begin{figure}[hbtp]
  \begin{center}
    \resizebox{0.8\textwidth}{!}{\includegraphics{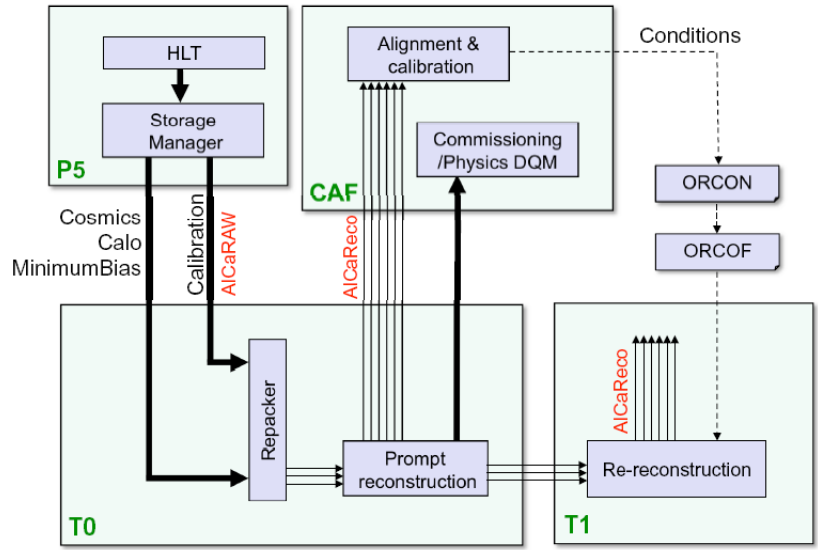}}
    \caption{Offline workflow for alignment and calibration used during CRAFT.}
    \label{fig:AlcaWorkflow}
  \end{center}
\end{figure}

The basic offline workflow for alignment and calibration in CRAFT was a
slightly simplified version of the full model for collisions, and it is
illustrated in Fig.~\ref{fig:AlcaWorkflow}. Commissioning experience
from this workflow in the context of a challenge with simulated events
has been reported elsewhere~\cite{csa08paper}. Event information
relevant for alignment and calibration was streamed from the CMS
detector site via the standard physics event stream 
(Section~\ref{sec:hlt}), and via a special calibration stream
and streams with special event content, labeled ``AlCaRaw'' (described below),
dedicated to particular calibration procedures.  Events from these
streams passed the conversion to the ROOT-based event data format at the Tier-0 (Section~\ref{sec:t0}) and in the case of the physics
stream entered the prompt reconstruction process. The reconstructed
data were then skimmed to create a series of ``AlCaReco''
datasets that were transferred to the CAF to be used as input to
alignment and calibration algorithms. The AlCaReco datasets are designed
to contain only the minimal amount of information required by the
associated alignment and calibration workflows. The skims producing
them performed both event selection, starting from a selection based
on HLT bits, and reduction of event content.  The alignment and
calibration workflows, performed at the CAF, used the AlCaReco
datasets to generate alignment and calibration constants that are
validated and uploaded to the conditions database. Re-reconstruction
at the Tier-1 sites, using the new constants, also generated
new AlCaReco datasets that were used in turn as input to the next
series of improvements on alignment and calibration constants.

\subsection{AlCaRaw streams}

Some calibration procedures in CMS require a very high event rate of
typically a few kHz in order to achieve the targeted precision in the
time scale of a few days. These events would saturate the available
bandwidth between the CMS cavern and the Tier-0 site if the full event
content were transferred. This concerns in particular the $\phi$ symmetry calibration procedures for ECAL and HCAL, and the ECAL
calibration with neutral pion decays. The solution is the creation of
special data streams called AlCaRaw already within dedicated high-rate
triggers at the HLT farm, which contain only the minimal information
needed for these workflows. These AlCaRaw streams have been
successfully generated for a significant part of the CRAFT run,
accumulating over 160 million events (for each, ECAL and HCAL) for
the $\phi$-symmetry streams.  Detailed information about the AlCaRaw
streams produced in CRAFT can be found in Ref.~\cite{CFT-09-020}.

\subsection{AlCaReco skims}
During the global runs that CMS performed in 2008, the number of AlCaReco
skims produced as part of prompt Tier-0 processing has been steadily
increased. The set of AlCaReco datasets that have been produced in
CRAFT is listed in Table~\ref{tab:alcarecoDS}. This list also contains
datasets that are not meaningful for calibration with cosmic muon data
and have been included only for commissioning the production
system. As a result, nine AlCaReco skims have been created
in parallel, which is comparable with the maximum number anticipated
for a given PD during LHC collisions, thus
constituting an important scaling test for the CMS alignment and
calibration framework. The number of events given in the table, which
corresponds to the output of the prompt processing, reflects the
selection mechanism. For example, the MuAlStandaloneCosmics dataset
did not require tracker information and thus selected a large part of
the overall CRAFT event sample. The TkAlCosmics0T dataset, which was
originally designed for runs without field, required a trajectory reconstructed in the tracker and
thus selected slightly more than one percent of the standalone muon
sample. The TkAlCosmicsHLT skim required the corresponding HLT trigger
bit and selected only particles with a transverse momentum cut above 4~GeV/c, which
resulted in a slightly smaller sample. Low noise thresholds allowed
the population of the HcalCalDiJets sample.

\begin{table}
\caption{AlCaReco datasets produced in CRAFT.}
\ \\
\begin{tabular}{|l|r|l|}
\hline
Dataset & Number of events & Purpose \\
\hline
\hline
TkAlCosmicsHLT & 4.3 M & Tracker alignment \\
\hline
TkAlCosmics0T  & 4.9 M & Tracker alignment (no ${\rm p_T}$ cut) \\
\hline
MuAlStandaloneCosmics & 288 M & Muon standalone alignment \\
\hline
MuAlBeamHaloOverlaps & 3.3 M & Muon endcap alignment \\
\hline
MuAlGlobalCosmics & 5.5 M & Muon system alignment w.r.t.\ tracker \\
\hline
HcalCalHOCosmics & 313 M & HCAL HO calibration \\
\hline
HcalCalDiJets & 67 M & HCAL calibration \\
\hline
MuAlCalIsolatedMu & 52 M & Muon system alignment, DT calibration \\
\hline
RpcCalHLT & 241 M & DT calibration, RPC monitoring \\
\hline
\end{tabular}
\label{tab:alcarecoDS}
\end{table}

\subsection{Alignment and calibration workflows, validation and
  sign-off of constants} All workflows deriving alignment and
calibration constants have been performed at the CAF based on the
AlCaReco datasets. The derived constants have been uploaded to the CMS
conditions database. The management of conditions within the database
is explained in more detail in
Section~\ref{sec:conditions}. Standardised validation procedures have
been applied to certify the correctness of the constants. The
validation results have been reviewed in a formalised sign-off
procedure, to ensure the quality of database constants that are
used for any central processing. Special care has been taken regarding
the interdependencies of the various workflows.

\subsection{Conditions used for reprocessing campaigns}
\label{sec:AlCaReprocessing}
The first comprehensive alignment and calibration campaign started
immediately after the end of data taking. The set of constants used
for the first reprocessing of the full CRAFT dataset included the
following: tracker alignment and alignment error estimates, strip
tracker gain calibration, bad strip and bad fiber maps, pixel tracker
gain and pedestal constants, internal track-based alignment constants
for the barrel muon DT chambers~\cite{CFT-09-016}, global longitudinal positions and tilt angles from the
optical alignment systems for most of the endcap muon chambers~\cite{CFT-09-017}, muon
DT inter channel synchronization, time pedestal and noise
calibration constants~\cite{CFT-09-023}, gain and pedestal calibration constants for
both ECAL and HCAL. All constants were ready, validated, signed-off
and included in the official set of conditions on 20 November 2008,
about two weeks after the end of data taking.

The second pass of alignment and calibration was performed after the
results of the first reprocessing became available around the middle of
January 2009. Tracker gain calibration was updated and
calibration of the Lorentz angle was added. Tracker alignment was
further improved~\cite{CFT-09-003}, benefiting also from the Lorentz angle calibration,
and from an increased number of pixel hits available due to the
updated alignment error estimates. The muon chambers were aligned relative
to the tracker with global tracks~\cite{CFT-09-016}; in addition, the optical alignment 
of the endcap system was extended~\cite{CFT-09-017}. Drift-tube calibration was updated following an
improved reconstruction in the inner chambers of the wheels closest to
the endcaps in the presence of the magnetic field~\cite{CFT-09-023}. HCAL pedestals and
gains were improved, and inter-calibration constants for the ECAL endcaps were updated based on 
laboratory measurements combined with information from laser data taken during CRAFT~\cite{CFT-09-004}. 
These constants were used in the second reprocessing of the CRAFT data.





\section{Conditions}
\label{sec:conditions}

The CMS conditions database system
relies on three databases for storing non-event data:
\begin{enumerate}
\item OMDS (Online Master Database System) is in the online network at the
detector site; it stores the data needed for the
configuration and proper settings of the detector ({\it{configuration
data}}), and the {\it{conditions data}} produced directly from the
front-end electronics. For example, the Data Control System (DCS)
information is stored with the ORACLE interface provided by
PVSS~\cite{PVSS}.
\item ORCON (Offline Reconstruction Condition DB Online subset) is also located at the 
detector site.  It stores all the condition data, including 
calibration and alignment constants, that are needed for the reconstruction of 
physics quantities in the HLT, as well as for detector performance studies.
These are a small subset of all the online constants.  These data are
written using the POOL-ORA~\cite{POOLORA} technology and are retrieved
by the HLT programs as {\texttt{C++}} objects.
\item ORCOFF (Offline Reconstruction Condition DB Offline subset) is located at the 
CERN computing centre.  It contains a copy of the information in ORCON, 
kept in sync through ORACLE streaming~\cite{Oracle}.
Data are retrieved by the reconstruction algorithms as
\texttt{C++} objects.
\end{enumerate}

In order to guarantee consistency of the data in ORCON and ORCOFF, it is one of the CMS 
policies to write any condition data needed for offline purposes to the ORCON
database. ORACLE streaming provides the transfer from ORCON to
ORCOFF.

\subsection{Interval of validity for conditions data and the global tag}

All conditions data are organised by condition database tags. A tag
points to one or more instances of a given type of condition data (e.g.\ 
ECAL pedestals), each of which has an associated interval of validity (IOV).  
The IOV is a range of events which is contiguous in time, for which that 
version of the condition data is valid. This range is normally defined in 
terms of run numbers, but can also be defined in terms of absolute time.  
While some conditions are only valid for the specific run for which they 
are measured (e.g.\ beamspot, pedestals), other conditions can be valid for any 
run (e.g.\ calorimeter intercalibration constants).  Each payload object in 
ORCON/ORCOFF is unambiguously indexed by its IOV and a tag.

The full consistent set of conditions which needs to be accessed by the HLT and 
offline reconstruction software is defined in a \emph{global tag}, which 
consists of one tag for each type of condition data. For a given event, 
the reconstruction algorithms query the corresponding conditions data by 
means of the global tag.

\subsection{Population of the conditions database}

\begin{figure}[hbtp]
  \begin{center}
    \resizebox{1.0\textwidth}{!}{\includegraphics{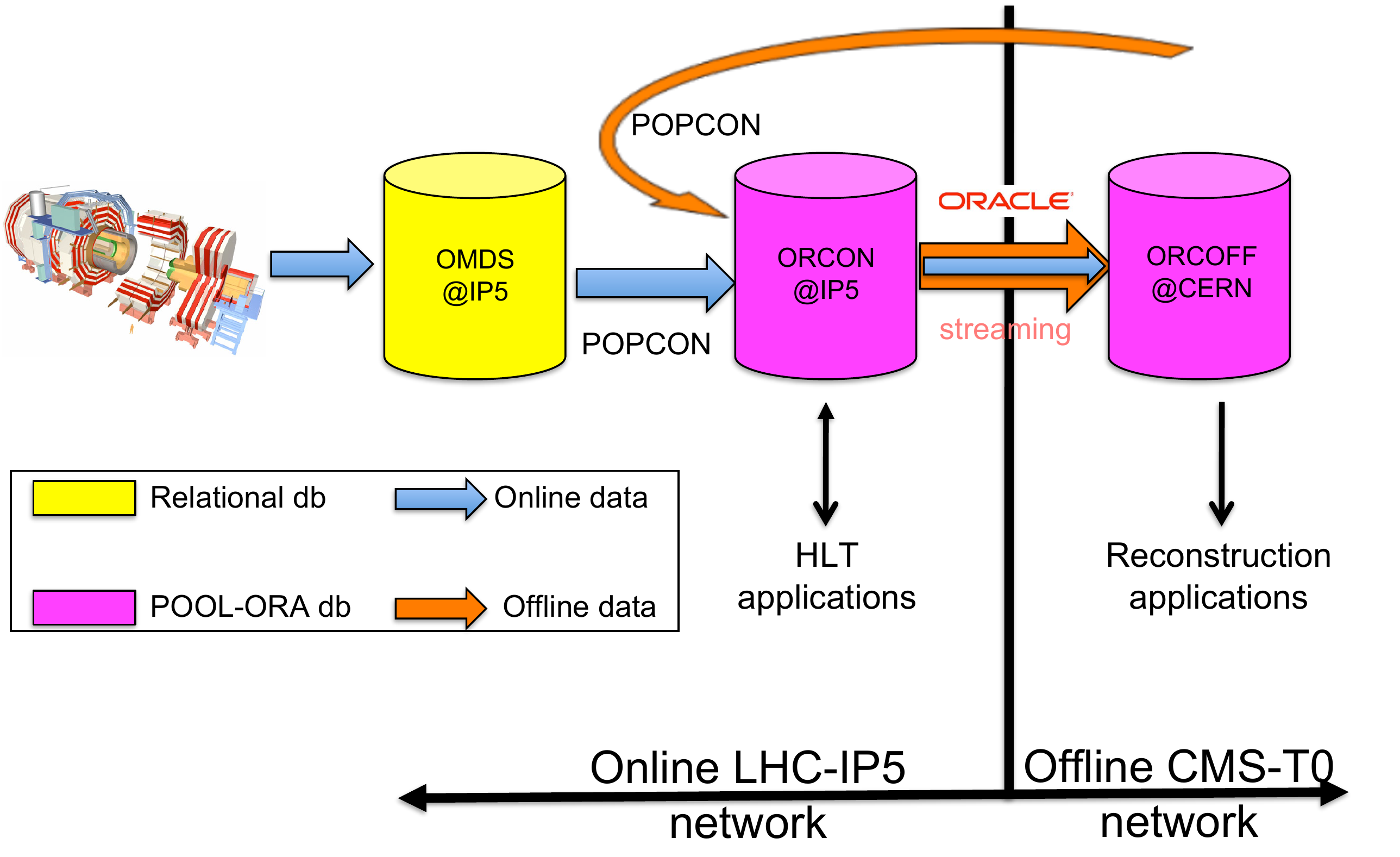}} \caption{Conditions databases architecture.}
    \label{fig:CondDBArchitecture}
  \end{center}
\end{figure}

The flow of conditions data is illustrated in 
Fig.~\ref{fig:CondDBArchitecture}. Conditions data that are produced online are initially stored 
in OMDS.  The subset of online conditions that are required for the HLT and 
offline reconstruction is extracted and sent to ORCON.  This data transfer is 
operated using a framework named PopCon (Populator of Condition
\cite{PopCon}).  PopCon encapsulates the relational data as POOL-ORA objects 
and adds meta-data information (the tag to which the object belongs and the 
IOV), so that the data is correctly indexed for reading by the HLT and offline 
software. Moreover, PopCon has the additional functionality of logging specific 
information about any transaction writing to the ORCON database.

Further conditions are produced by the alignment and calibration
workflows operated offline, as described in
Section~\ref{sec:alcareco}; these are directly uploaded to the
ORCON database, again using PopCon. 

Finally, all data in ORCON are transferred to ORCOFF, which is the database 
used for all offline processing and analysis, via ORACLE streaming.

For massively parallel read-access, the ORCON and ORCOFF databases are
interfaced with a cache system referred to as ``FroNTier,'' which in the case of ORCOFF is the mechanism used to distribute conditions data
to the Tier-1 and Tier-2 centres outside CERN. Caching servers are 
used to cache requested objects to avoid repeated access to the same data, 
thus significantly improving the performance and greatly reducing the load on
the central database servers. Further details can be found
in Ref.~\cite{FRONTIER}.

\subsection{Database population during CRAFT}\label{sec:CondDbPopIn2008}

During the CRAFT run, the majority of conditions data
were transferred in the offline database using the PopCon application.
A central procedure, based on an automatic uploader via a dedicated machine 
in the online network, was successfully deployed during 2008 \cite{PopCon}.

A set of automatic jobs was set up for each sub-detector, in 
order to both populate the ORCON database and monitor any transaction to it.
Each automatic job is associated with a ``watchdog'' tool that
monitors its status.  A dedicated web interface was set up on a CMS web 
server in order to monitor all database transactions. PopCon was used by almost all sub-detectors and an average
of one hundred PopCon applications per day were run during CRAFT.

During the entire duration of CRAFT the total amount of conditions data 
written to ORCON was about 1 TB. ORCON-ORCOFF streaming and the FroNTier 
caching mechanism operated smoothly throughout CRAFT.




\section{Analysis Model and Tool}
\label{sec:analysis}
In this section, the model and the tool to analyse the recorded and
reconstructed data are described. CMS uses a distributed data-analysis
model~\cite{CompModel} mostly based on the WLCG Grid
infrastructure. It also supports low-latency access to 
data on the CAF for prompt analysis and calibration.  The CMS analysis model is
data-location driven, i.e. the user analysis runs where data are
located.  The related workflow is mainly characterised by the
following steps: interactive code development using small data
samples; job preparation and configuration to run over higher
statistics (hence to access the whole dataset or a significant part of
it); and interactive analysis of the obtained results.  With the
increasing complexity of the computing infrastructure, the
implementation of such a workflow became more and more difficult
for the end user.  In order to provide the physicists an
efficient access to the distributed data while hiding the underlying
complexity, CMS developed and deployed a dedicated tool named
 CMS Remote Analysis
Builder (CRAB)~\cite{Crab}, which is described in the following.

\subsection{CRAB architecture}
\label{sec:CrabArchitecture}  
CRAB is the official CMS tool for distributed analysis.  The system,
which guarantees interoperability with the various grid flavours and
batch submission systems, has evolved into a client-server
architecture. The client provides the
user interface and is a command line
Python~\cite{Python} application which mainly takes care of the local
environment interaction and packages private user library and code, in order
to replicate remotely the very same local configuration. The server
is the intermediate service responsible to automate
the user analysis workflow with resubmissions, error handling, and output retrieval
thus leaving to the user just the preparation of the configuration
file. The server also notifies the user of the output availability.
The server architecture is made of a set of independent components
communicating asynchronously through a shared messaging service and cooperating
to carry out the analysis workflow.

\begin{figure}[b!]
  \begin{center}
   \includegraphics[height=4.8cm,width=7.5cm]{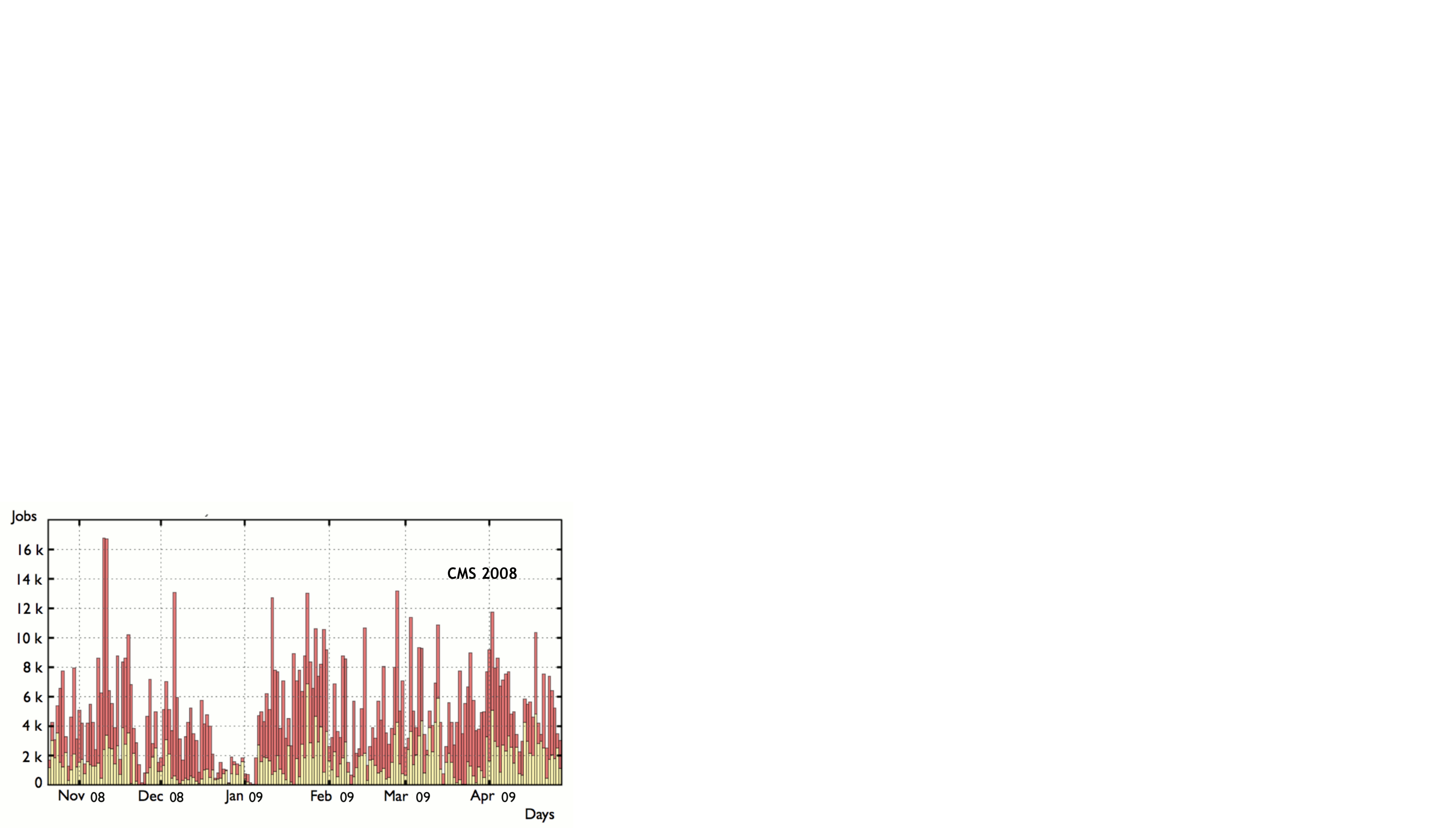}
   \includegraphics[height=4.8cm,width=7.5cm]{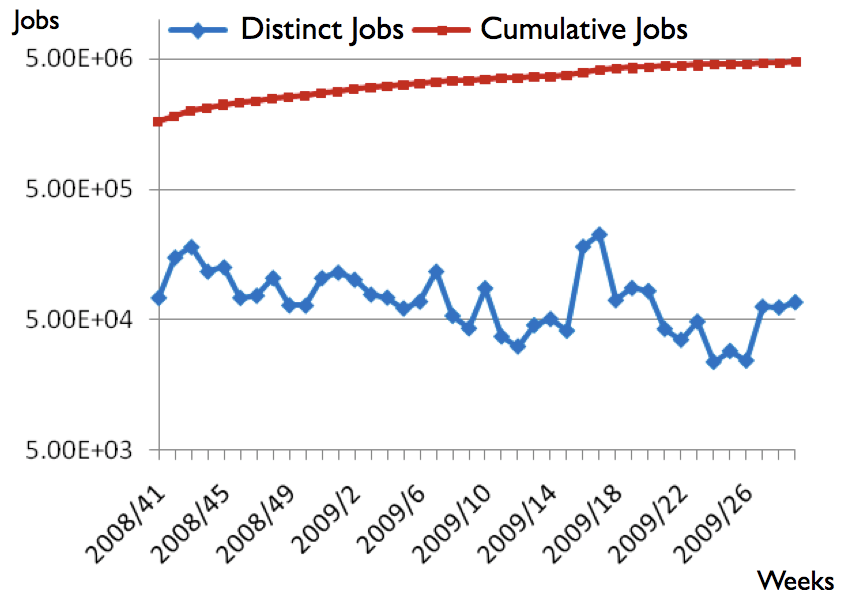}
\caption{CRAFT jobs distributions as a function of time. Left: Daily
         distribution of analysis jobs submitted using CRAB and
         accessing CRAFT data.  Grid (dark shading, red) and CAF (light shading, yellow) activities
         are shown.  (Taken from monitoring sources). Right: CRAFT jobs submitted only at CAF (with and
         without CRAB). The upper line shows the cumulative number of jobs, the lower 
line shows the number of jobs submitted each week. The time window extends
         well beyond the end of CRAFT data taking to cover the
         extensive period of analysis.}
\label{fig:job}
  \end{center}
\end{figure}

\subsection{CRAFT analysis activity}
\label{sec:AnaysisActivity}  
The CRAFT data have been analysed both at CERN (using the local batch
system at the CAF), and on the Grid making use of distributed resources
(Tier-2).  While access to data on Tier-2 sites has been performed exclusively by CRAB,
the CAF queues have been used to run both CRAB and non-CRAB jobs.  The
large fraction of non-CRAB jobs executed at the CAF is partially due
to calibration and alignment workflows, which are for the time being
not integrated within the CRAB framework. The collaboration is
currently evaluating the best strategy for a proper
integration of such workflows.

\begin{figure}[t!]
\begin{center}
\includegraphics[height=4.7cm]{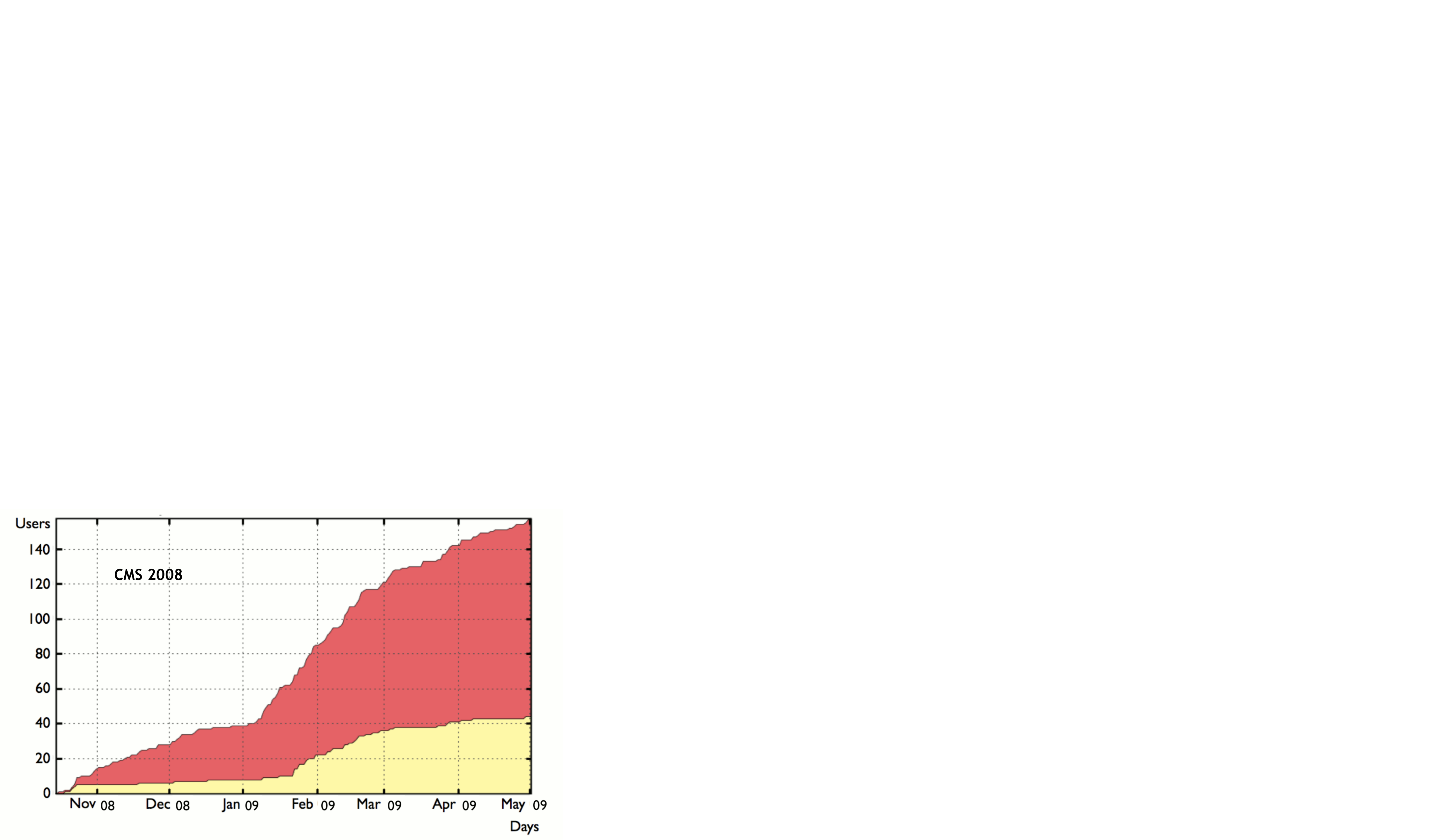}
\includegraphics[height=4.7cm]{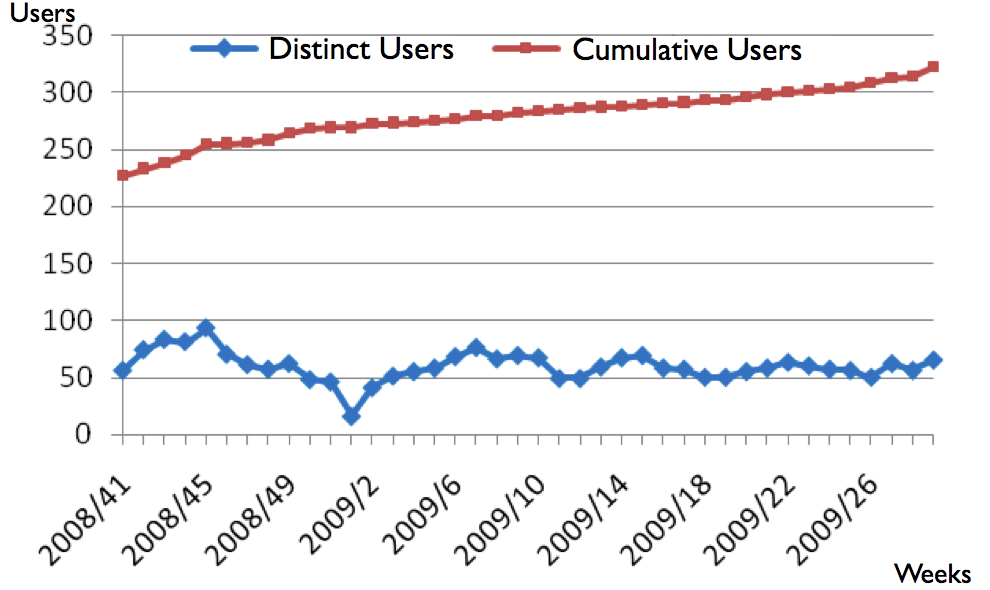}
\caption{Cumulative plot of number of different users accessing CRAFT
         data as a function of time. Left: users using CRAB to submit
         Grid (dark shading, red) and CAF (light shading, yellow) jobs.  (Taken from monitoring sources). Right: number 
         of users submitting jobs
         only at CAF (with and without CRAB). The lower line shows the 
         number of users per week, the upper line the 
         integrated number over a long period.
         The time window extends
         well beyond the end of CRAFT data taking to cover the
         extensive period of analysis.}
\label{fig:user}
\end{center}
\end{figure}

\subsection{Analysed data volume}
\label{sec:DataVolume}  
\begin{figure}[h!]
\begin{center}
   \resizebox{8.5cm}{!}{\includegraphics{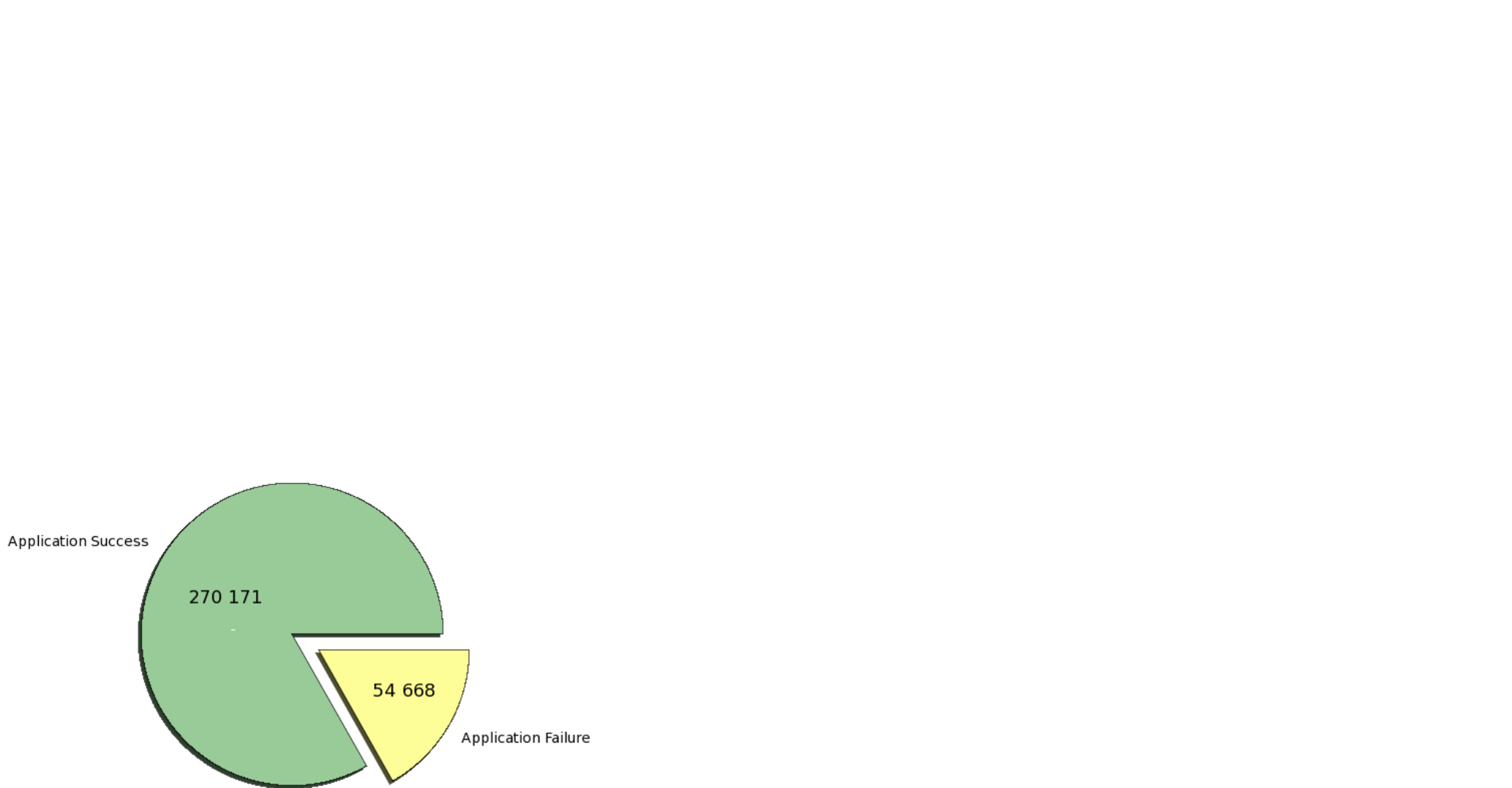}}
   \resizebox{6.7cm}{!}{\includegraphics{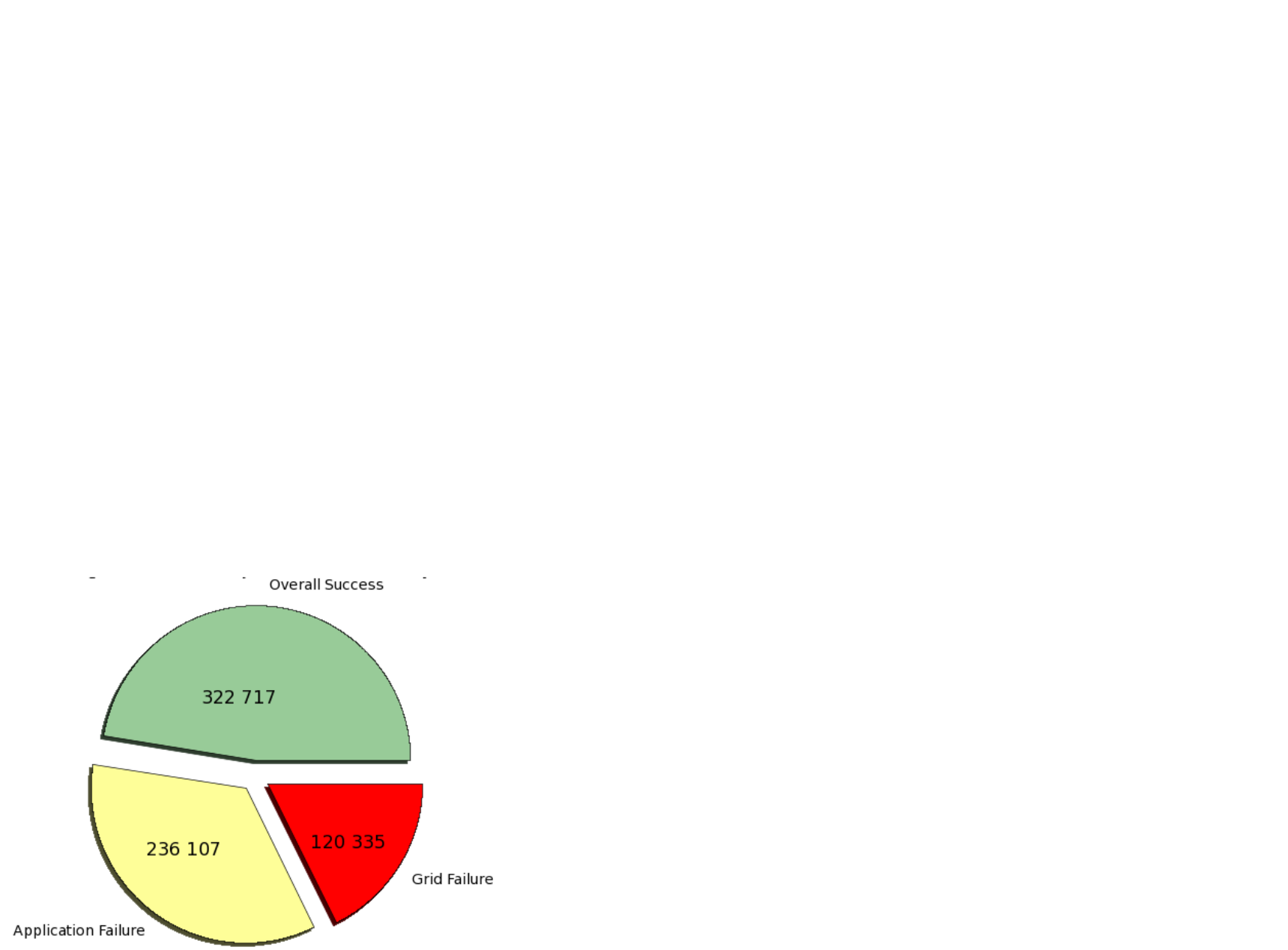}}
\caption{Success rate of CRAFT analysis jobs submitted using CRAB. Left: jobs submitted only at CAF. Right: jobs submitted through the Grid.} 
\label{fig:efficiency}
\end{center}
\end{figure}

From October 2008 to the beginning of May 2009 more than 2 million
analysis jobs accessed CRAFT data, including both CRAB and non-CRAB
jobs. The quoted value takes into account both CAF and Grid activity
(Fig.~\ref{fig:job}). Figure~\ref{fig:user} shows the
cumulative numbers of distinct users which performed CRAFT data
analysis in the considered time window. The shapes, combined with
daily jobs distribution, give a clear indication of how the user
community increased continuously. Referring to the same time interval
it is estimated that more than 200 distinct users in total performed
CRAFT analysis activities.  As shown in Fig.~\ref{fig:efficiency}, the
overall efficiency of CRAFT analysis jobs is approximately 60\%. Local
submissions on the CAF were 85\% efficient.  The main source of
failures of Grid CRAFT jobs are remote stage-out
problems, which will be addressed in the future by a new workload
management infrastructure. In general, there is a 10\% failure rate due to problems within the user code. No relevant
bottlenecks were experienced by the system during CRAFT.

\section{Summary}
\label{sec:summary}
Data taking with cosmic ray muons during the CRAFT exercise in 2008, which
lasted about a month, has provided a wealth of experience in operating
the workflows from recording to analysing the data. The online system and
the high level trigger have been operated continuously, and besides
stress-testing the general reliability, major functionalities have been
exercised. These include the definition of streams and primary
datasets, and the interaction with associated HLT menus, for which
efficient online-offline interplay is essential.

Data handling has been confronted with almost the full qualitative
complexity expected for collisions. Most of the Tier-0 related
processing has been handled with the final software infrastructure and
performed very well. The setup for collisions will still require
the ramp-up of the express stream infrastructure for prompt
calibration and monitoring, and inclusion of alignment and calibration
skims into the Tier-0 processing system. Data distribution via
the PhEDEx system performed very well overall. Re-reconstruction at the Tier-1 sites was performed with adequate
turn-around time, but showed the need of a system with full accountability,
which will be introduced by spring 2010.

Event reconstruction has used various algorithms dedicated to
cosmic ray muons, but in addition used CRAFT to commission the methodology
for collisions. A comprehensive set of reconstructed objects
has been provided to support the analysis of the CRAFT
data. The workflow for fast deployment of code corrections presented some organisational challenges, and while solutions were generally
available quickly, several improvements for future operation were
implemented. Data quality monitoring was performed both at the online and
offline levels, and regular DQM shifts were run continuously. Remote CMS centres fully participated in this
process. Further certification and validation of the data were
performed by prompt feedback groups who analysed the output of the
prompt reconstruction, and discovered time-dependent developments
which were correlated to intentional changes in the detector
conditions.

Essentially all alignment and calibration constants accessible with
cosmic ray muon data taking have been determined during CRAFT, thus
putting the corresponding framework through a very comprehensive test. The
associated organizational challenges with a large number of concurrent
workflows, properly respecting the interdependencies, were successfully addressed. Several reprocessing campaigns with successively improved
constants have been performed, which provided a very high data-quality
sample for cosmic ray analysis. The conditions database, which is a sophisticated
network of commercial relational database management servers with
proxies and mechanisms to provide distributed access, proved to be a solid
basis for all conditions-related operations.

The large sample of cosmic ray data also provided a realistic test of
the distributed analysis system. Limitations in job execution
efficiency were traced to remote file staging issues, which will
be addressed by future improvements in the workload management
system. Overall, CRAFT has shown that CMS has highly reliable methods at its
disposal to make data samples available with short latency for analysis at remote centres.

In conclusion, CRAFT has demonstrated the proper functioning of the overall
CMS workflow machinery to a very high degree. While the challenge has
been instrumental in identifying individual areas which need some further
improvements, the overall system is well designed and is
expected to scale smoothly to data taking with LHC collisions.

\section*{Acknowledgements}
We thank the technical and administrative staff at CERN and other CMS
Institutes, and acknowledge support from: FMSR (Austria); FNRS and FWO
(Belgium); CNPq, CAPES, FAPERJ, and FAPESP (Brazil); MES (Bulgaria);
CERN; CAS, MoST, and NSFC (China); COLCIENCIAS (Colombia); MSES
(Croatia); RPF (Cyprus); Academy of Sciences and NICPB (Estonia);
Academy of Finland, ME, and HIP (Finland); CEA and CNRS/IN2P3
(France); BMBF, DFG, and HGF (Germany); GSRT (Greece); OTKA and NKTH
(Hungary); DAE and DST (India); IPM (Iran); SFI (Ireland); INFN
(Italy); NRF (Korea); LAS (Lithuania); CINVESTAV, CONACYT, SEP, and
UASLP-FAI (Mexico); PAEC (Pakistan); SCSR (Poland); FCT (Portugal);
JINR (Armenia, Belarus, Georgia, Ukraine, Uzbekistan); MST and MAE
(Russia); MSTDS (Serbia); MICINN and CPAN (Spain); Swiss Funding
Agencies (Switzerland); NSC (Taipei); TUBITAK and TAEK (Turkey); STFC
(United Kingdom); DOE and NSF (USA). Individuals have received support
from the Marie-Curie IEF program (European Union); the Leventis
Foundation; the A. P. Sloan Foundation; and the Alexander von Humboldt
Foundation.


\bibliography{auto_generated} 

\cleardoublepage\appendix\section{The CMS Collaboration \label{app:collab}}\begin{sloppypar}\hyphenpenalty=500\textbf{Yerevan Physics Institute,  Yerevan,  Armenia}\\*[0pt]
S.~Chatrchyan, V.~Khachatryan, A.M.~Sirunyan
\vskip\cmsinstskip
\textbf{Institut f\"{u}r Hochenergiephysik der OeAW,  Wien,  Austria}\\*[0pt]
W.~Adam, B.~Arnold, H.~Bergauer, T.~Bergauer, M.~Dragicevic, M.~Eichberger, J.~Er\"{o}
% ...
\vskip\cmsinstskip
% ...

\end{sloppypar}
\end{document}